\documentclass[aps, prb, twocolumn, superscriptaddress, showkeys]{revtex4-2}
      
\usepackage[utf8]{inputenc}
\usepackage[T1]{fontenc}
\usepackage{dcolumn}
\usepackage{bm}					
\usepackage{graphicx}
\usepackage{amsmath, amsfonts, amssymb, mathtools}	
\usepackage{latexsym} 				
\usepackage{xcolor}
\usepackage[caption = false]{subfig}
\usepackage[breaklinks, linkcolor = black]{hyperref}
\usepackage{tikz}
\usepackage{floatrow}
\usepackage{multirow}
\usepackage{makecell}
\usepackage{braket}
\usepackage{diagbox}
\usepackage{ulem} 
\usepackage{textcomp}
\usepackage{soul}
\usepackage{booktabs} 
\usepackage{siunitx}

\DeclareMathOperator\erf{erf}

\begin{document}

\title{Electronic and Optical Excitations in van der Waals Materials from a Non-Empirical Wannier-Localized Optimally-Tuned Screened Range-Separated Hybrid Functional}

%
\author{Mar\'ia \surname{Camarasa-G\'{o}mez}}
\email{maria.camarasa-gomez@weizmann.ac.il}
\affiliation{Department of Molecular Chemistry and Materials Science, Weizmann Institute of Science, Rehovoth 7610001, Israel}

\author{Stephen E. Gant}
\affiliation{Department of Physics, University of California, Berkeley, CA 94720, USA}
\affiliation{Materials Sciences Division, Lawrence Berkeley National Laboratory, Berkeley, CA 94720, USA}

\author{Guy Ohad}
\affiliation{Department of Molecular Chemistry and Materials Science, Weizmann Institute of Science, Rehovoth 7610001, Israel}

\author{Jeffrey B. Neaton}
\affiliation{Department of Physics, University of California, Berkeley, CA 94720, USA}
\affiliation{Materials Sciences Division, Lawrence Berkeley National Laboratory, Berkeley, CA 94720, USA}
\affiliation{Kavli Energy NanoSciences Institute at Berkeley, University of California, Berkeley, CA 94720, USA}

\author{Ashwin Ramasubramaniam}
\email{ashwin@umass.edu}
\affiliation{Department of Mechanical and Industrial Engineering, University of Massachusetts Amherst, Amherst MA 01003, USA}
\affiliation{Materials Science and Engineering Graduate Program, University of Massachusetts, Amherst, Amherst MA 01003, USA}

\author{Leeor Kronik}
\email{leeor.kronik@weizmann.ac.il}
\affiliation{Department of Molecular Chemistry and Materials Science, Weizmann Institute of Science, Rehovoth 7610001, Israel}


\date{\today}

\begin{abstract}
Accurate prediction of electronic and optical excitations in van der Waals (vdW) materials is a long-standing challenge for density functional theory. The recently proposed Wannier-localized optimally-tuned screened range-separated hybrid (WOT-SRSH) functional has proven successful in non-empirical determination of electronic band gaps and optical absorption spectra for various covalent and ionic crystals. However, for vdW materials the tuning of the material- and structure-dependent functional parameters has, until now, only been attained semi-empirically. Here, we present a non-empirical WOT-SRSH approach applicable to vdW materials, with the optimal functional parameters transferable between monolayer and bulk. We apply this methodology to prototypical vdW materials: black phosphrous, moldybdenum disulfide, and hexagonal boron nitride (in the latter case including zero-point renormalization). We show that the WOT-SRSH approach consistently achieves accuracy levels comparable to experiments and \textit{ab initio} many-body perturbation theory (MBPT) calculations for band structures and optical absorption spectra, both on its own and as an optimal starting point for MBPT calculations. 

\end{abstract} 
 
\maketitle

Van der Waals (vdW) materials \cite{Geim2004, Geim2005, Wang2012, Geim2013, Liu2016, Novoselov2016, Ajayan2016}, comprised of weakly interacting stacks of (quasi-)two-dimensional (2D) layers, have attracted much interest owing to the outstanding tunability of their electronic and optical properties with the number, composition, and orientation of individual 2D layers.
The unique properties of these materials have also motivated an ongoing effort in accurate prediction of their electronic band structures and optical absorption spectra \cite{Bernardi2017,Yadav2023}. Often this is based on \textit{ab initio} many-body perturbation theory (MBPT) \cite{Hedin1965,Onida2002} within the framework of the GW method \cite{Hybertsen1986} and the Bethe-Salpeter equation (BSE) \cite{Rohlfing1998, Onida1998}. However, there is significant interest in more readily affordable computational approaches rooted in density functional theory (DFT) \cite{Parr1989, Dreizler1990}.
Recent methodological developments within DFT have greatly improved the quantitative accuracy of band gap predictions in solids. Notable examples are 
Koopmans-compliant functionals \cite{Ma2016, Weng2017, Nguyen2018, Colonna2019, Colonna2022, Linscott2023}, localized orbital scaling corrections \cite{Li2018, Mei2020, Mahler2022}, dielectric-dependent functionals \cite{Shimazaki2009, Skone2014, Skone2016, Chen2018, Sun2020, Galli2023}, and screened range-separated hybrid (SRSH) functionals \cite{Refaely2013, Refaely2015,Miceli2018,Bischoff2019,Yang2023,Wing2021}.

Here, we focus on a recently introduced specific class of SRSH functionals, Wannier-localized optimally-tuned (WOT)-SRSH functionals \cite{Wing2021, Ohad2022, Gant2022, Ohad2023}, because they combine three advantages: They are non-empirical, with physical parameters derived directly from the pristine system \cite{Wing2021}; They can be applied to both molecules \cite{Ohad2024} and solids \cite{Wing2021}; They lend themselves naturally to computation of optical absorption directly from time-dependent (TD) DFT \cite{Refaely2015, Ohad2023,Kronik2016}. 

In the SRSH approach, correct asymptotic dielectric screening is introduced via screened long-range exchange, while short-range exchange provides a good balance with dynamic correlation and mitigates self-interaction errors \cite{Refaely2013, Kronik2016, Kronik2018}. The transition between short- and long-range exchange is governed by a range-separation parameter. For bulk solids, the fraction of short-range exchange and the range-separation parameter have been successfully determined non-empirically via the Wannier-localized optimal tuning procedure \cite{Wing2021}, elaborated below. For vdW materials, however, the need to address both monolayer and bulk with a consistent set of parameters has to date hindered use of the WOT approach. In prior work, the requisite SRSH parameters were determined semi-empirically by fitting to MBPT-computed gaps \cite{Ramasubramaniam2019, Camarasa-Gomez2023}, thereby strongly limiting the predictive power of the method for this important class of materials. 

In this article, we obtain non-empirical SRSH functionals suitable for vdW materials, entirely from first principles, via a generalized WOT procedure. Specifically, we present an internally self-consistent framework that exploits an ionization potential (IP) ansatz \cite{Ma2016, Wing2021} to determine optimal parameters of a fully non-empirical SRSH, for both monolayer and bulk phases of any specific vdW material. We apply this approach to three representative vdW materials -- black phosphorus (bP), molybdenum disulfide (MoS$_2$), and hexagonal-boron nitride (h-BN) -- that are prototypical examples of narrow, moderate, and wide gap semiconductors, respectively. Our calculated electronic bandstructures and optical absorption spectra are in excellent agreement with both experimental and MBPT studies for all materials. Even the gap of monolayer h-BN, which has previously been shown to be challenging for SRSH functionals~\cite{Ramasubramaniam2019}, is now predicted correctly once zero-point renormalization (ZPR) of the electronic band gap is properly taken into account. This confirms the accuracy of the self-consistent tuning approach and opens the door to truly predictive use of WOT-SRSH functionals in, e.g., high-throughput screening and design of vdW materials \cite{Haastrup2018, Zhang2019}.

\begin{figure*}[h!]
 \vspace{0.0cm}
    \includegraphics[width=0.85\linewidth]{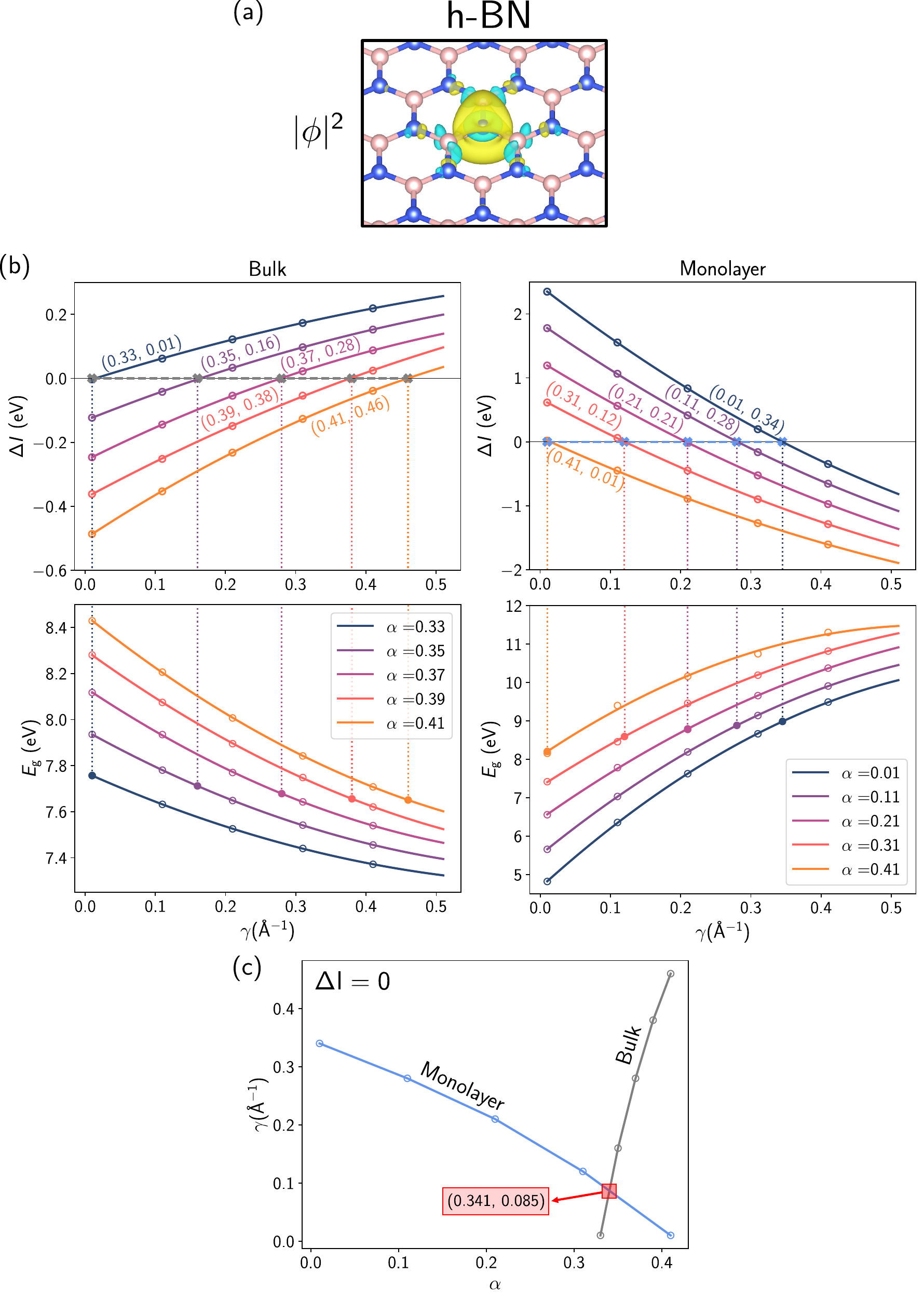}  \caption{Demonstration of the non-empirical WOT-SRSH method, as applied to h-BN: (a) highest-expectation-value occupied maximally-localized Wannier function; (b) The IP ansatz target function, $\Delta I$ (top) and the fundamental band gap, $E_g$, at the $K$-point (bottom), for both bulk ($\epsilon_\infty$ of the solid, left) and monolayer ($\epsilon_\infty = 1$, right), as a function of the range-separation parameter, $\gamma$, for different values of the fraction of short-range exact exchange, $\alpha$. Circles indicated computed data points and the lines are a guide to the eye. Closed circles have been obtained from $(\alpha, \gamma)$ pairs that obey the IP ansatz, with the parameter values shown; (c) IP ansatz fulfillment ($\Delta I = 0$) curves in the $(\alpha, \gamma)$ plane for  monolayer and bulk, with their point of intersection, $(\alpha^*, \gamma^*)$, obeying the IP ansatz for both phases simultaneously and used for predictive calculations.}\label{fig:method-IP-vdW} 
\end{figure*}

The SRSH functional is based on a decomposition of the Coulomb interaction, $1/r$ \cite{Leininger1997}, in the form
\cite{Yanai2004} 
\begin{equation}\label{eq:coulomb}
    \frac{1}{r} = \frac{\alpha + \beta \erf (\gamma r)}{r} + \frac{1-[\alpha + \beta \erf (\gamma r)]}{r},
\end{equation}
where $\erf(\cdot)$ is the error function and $\alpha$, $\beta$, $\gamma$ are parameters. The first term of Eq.\ (\ref{eq:coulomb}) is treated using exact (Fock) exchange (XX) whereas the second term is treated using semilocal (SL) DFT exchange. This naturally partitions the exchange interaction into short-range (SR) and long-range (LR) components, where the XX fraction is $\alpha$ in the SR and $\alpha+\beta$ in the LR (with complementary SL exchange, SLx, fractions), and $\gamma$ is the range-separation parameter. 

To enforce asymptotic screening of the Coulomb repulsion in a three-dimensional (3D) solid (neglecting anisotropy), we set $\alpha + \beta$ to  $1/{\epsilon_{\infty}}$, the high-frequency scalar dielectric constant (obtained as the average of the diagonal terms of the dielectric tensor) \cite{Refaely2013, Wing2019}. The corresponding non-multiplicative exchange potential, $v_x^{\textnormal{SRSH}}$, obtained within generalized Kohn-Sham theory \cite{Seidl1996, Kummel2008, Kronik2012, Perdew2017, Baer2018}, is then given by 
\begin{equation}
v_x^{\textnormal{SRSH}} = \alpha v_{\textnormal{XX}}^{\textnormal{SR}} + (1-\alpha)v_{\textnormal{SLx}}^{\textnormal{SR}}+\frac{1}{\epsilon_{\infty}}v_{\textnormal{XX}}^{\textnormal{LR}}+\left(1-\frac{1}{\epsilon_{\infty}} \right) v_{\textnormal{SLx}}^{\textnormal{LR}}.
\end{equation}
For 2D solids, we set $\epsilon_{\infty}=1$, as this is the formal exact asymptotic limit for screening in 2D materials \cite{Cudazzo2011, Andersen2015, Qiu2016}. While this is the correct asymptotic limit in all three directions, anisotropy in the division between short range and long range is still not accounted for explicitly, but we show below that this is already sufficient for obtaining excellent results nonetheless.

While $\epsilon_\infty$ can be calculated routinely from first principles \cite{Nunes2001,Souza2002}, $\alpha$ and $\gamma$ must be determined by other means. Previously, this has only been accomplished by simultaneous fitting to benchmark band gaps of monolayer and bulk phases~\cite{Ramasubramaniam2019, Camarasa-Gomez2023}, an approach denoted henceforth as semi-empirical (SE) SRSH. Here, we move beyond it by generalizing the non-empirical WOT scheme,\cite{Wing2021, Ohad2022, Ohad2023, Gant2022} previously used in 3D solids. WOT enforces the IP ansatz for removal of charge from the maximally localized Wannier function, $\phi$, with the highest one-electron expectation energy, such that for any given value of $\alpha$ we seek $\gamma$ that obeys 
\begin{equation}
\label{eq:deltaI}
  \Delta I^\gamma=  E_{\textrm{constr}}^\gamma[\phi](N-1)- E^\gamma(N)+ \bra{\phi} \hat{H}_{\textrm{SRSH}}^\gamma \ket{\phi} = 0,
\end{equation}
where $E^\gamma(N)$ is the total energy of the neutral $N$-electron system, $E_{\textrm{constr}}^\gamma[\phi](N-1)$ is the total energy (calculated via a constrained minimization procedure \cite{Wing2021}) of the system with one electron subtracted from $\phi$, and $\bra{\phi} \hat{H}_{\textrm{SRSH}}^\gamma \ket{\phi}$ is the expectation value of the SRSH Hamiltonian, $\hat{H}_{\textrm{SRSH}}^\gamma$, with respect to the Wannier function. An example of $\phi$, for h-BN, is given in Fig.\ \ref{fig:method-IP-vdW}(a) [Section I of the Supporting Information (SI) contains the equivalent of Fig.\ \ref{fig:method-IP-vdW} for black phosphorus (bP) and molybdenum disulfide (MoS$_2$)]. 

\begin{table*}
\footnotesize
\centering
\setlength{\tabcolsep}{3.0pt} 
	\begin{tabular}{c|ccccccccc}
	\toprule
		Material	& Phase	& $E_g^\textnormal{GW@SE}$ [eV] & $E_g^\textnormal{GW@WOT}$ [eV] & $E_g^\textnormal{SE}$ [eV] & $E_g^\textnormal{WOT}$ [eV] & $E_g^{\textnormal{lit.}}$ [eV] \\
		\noalign{\vskip 0.1cm}    
		\hline
		 \noalign{\vskip 0.1cm}    
\noalign{\vskip 0.15cm} 
\multirow{2}{*}{\textbf{bP}} & Bulk &  $0.50$ & $0.47$  & $0.56$  &  $0.40$ & 0.58\textsuperscript{A}, 0.3\textsuperscript{B}\\
& 1L & $2.04$ &  $1.98$ & $1.95$ & $1.77$ & 2.00\textsuperscript{B} (2.03\textsuperscript{C})\\
\noalign{\vskip 0.15cm} 
\multirow{2}{*}{\textbf{MoS\textsubscript{2}}} & Bulk &  $2.12$ & $2.13$  & $2.03$  &  $2.11$ & (2.00\textsuperscript{D})\\
& 1L	& $2.67$ ($2.61$) &  $2.54$ ($2.60$) & $2.65$ ($2.55$) & $2.61$ ($2.51$) & 2.47\textsuperscript{E}, (2.78\textsuperscript{F}, 2.53\textsuperscript{C})\\
\noalign{\vskip 0.15cm} 
\multirow{2}{*}{\textbf{h-BN}} & Bulk	&  $7.36$ [$7.10$] & $7.77$ [$7.51$]  & $6.63$ [$6.37$] &  $7.78$ [$7.52$] & 6.99\textsuperscript{G}\\
& 1L  & $8.26$ [$7.88$] &  $8.70$ [$8.32$] & $7.26$ [$6.88$] & $8.51$ [$8.13$] & 8.14\textsuperscript{H}\\
		\bottomrule
	\end{tabular}
	\caption{GW band gaps based on SE-SRSH ($E_g^\textnormal{GW@SE}$) and  WOT-SRSH ($E_g^\textnormal{GW@WOT}$), as well as SE-SRSH ($E_g^\textnormal{SE}$) \cite{Ramasubramaniam2019, Camarasa-Gomez2023} and WOT-SRSH band gaps ($E_g^\textnormal{WOT}$) for the materials studied in this article, compared to computational literature band gaps ($E_g^{\textnormal{lit.}}$). All band gaps are direct, obtained at the $\Gamma$ point for black phosphorus and at the $K$ point for MoS$_2$ \protect{\cite{note1}} and h-BN. Numbers in parentheses correspond to calculations with spin-orbit coupling for MoS$_2$. Numbers in squared parentheses for h-BN correspond to band gaps after subtraction of a ZPR correction (0.26 eV for bulk, 0.38 for 1L).\\
    \textsuperscript{A}Ref.\ \onlinecite{Wang2015}, value from GW$_0$.
    \textsuperscript{B}Ref.\  \onlinecite{Tran2014}, value from G$_0$W$_0$.
    \textsuperscript{C}Refs.\  \onlinecite{Haastrup2018, Gjerding2021}, values from G$_0$W$_0$.
    \textsuperscript{D}Ref.\ \onlinecite{Komsa2012}, value from GW$_0$.  
    \textsuperscript{E}Ref.\  \onlinecite{Wilhelm2023}, value from G$_0$W$_0$.
    \textsuperscript{F}Ref.\  \onlinecite{Qiu2013}, value from G$_0$W$_0$.
    \textsuperscript{G}Ref.\ \onlinecite{Kolos2019}, value from G$_0$W$_0$. %
    \textsuperscript{H}Ref. \onlinecite{Louie2022}, value from eigenvalue-self-consistent GW$_0$. 
	}\label{t1}
\end{table*}

\begin{table*}
\footnotesize
\centering
\setlength{\tabcolsep}{3.0pt} %
	\begin{tabular}{c|ccccccccc}
	\toprule
		Material	& Phase	& $E_\textnormal{opt}^\textnormal{BSE@SE}$ [eV] & $E_\textnormal{opt}^\textnormal{BSE@WOT}$ [eV] & $E_\textnormal{opt}^\textnormal{TD-SE}$ [eV] & $E_\textnormal{opt}^\textnormal{TD-WOT}$ [eV] & $E_\textnormal{opt}^{\textnormal{lit.}}$ [eV] \\
		\noalign{\vskip 0.1cm}    
		\hline
		 \noalign{\vskip 0.1cm}    
\noalign{\vskip 0.15cm} 
\multirow{2}{*}{\textbf{bP}} & Bulk &  $0.39$ & $0.37$  & $0.32$  &  $0.26$ & 0.25\textsuperscript{A}\\
& 1L & $1.44$ &  $1.40$ & $1.37$ & $1.28$ & 1.20\textsuperscript{A} (1.45\textsuperscript{B})\\
\noalign{\vskip 0.15cm} 
\multirow{3}{*}{\textbf{MoS\textsubscript{2}}} & Bulk &  $1.98$ & $1.99$  & $1.94$  &  $1.99$ & ($1.88$\textsuperscript{C}) \\
& 1L	&\makecell{ $2.09$ \\ (A: 2.03,  B: 2.20)} & \makecell{ $2.00$ \\ (A: 2.03,  B: 2.18)} & \makecell{ $2.03$ \\ (A: 1.94,  B: 2.16)} & \makecell{ $2.06$ \\ (A: 1.97, B: 2.18)} & \makecell{ 2.12\textsuperscript{D} \\ (2.07\textsuperscript{B})}\\
\noalign{\vskip 0.15cm} 
\multirow{2}{*}{\textbf{h-BN}} & Bulk	&  $6.00\ [5.74]$ & $6.25\ [5.99]$  & $5.83\ [5.57]$  & $6.52\ [6.26]$  & 5.99\textsuperscript{E}\\
& 1L  & $5.96\ [5.58]$ &  $6.22\ [5.84]$ & $5.94\ [5.56]$ & $6.57\ [6.19] $ & 5.95\textsuperscript{F}, 6.03\textsuperscript{G}, 6.3\textsuperscript{H}\\
		\bottomrule
	\end{tabular}
 \caption{Optical gaps obtained from GW-BSE  
 and TD-SRSH ($E^\textnormal{GW-BSE}_\textnormal{opt}$),  ($E^\textnormal{TD-}_\textnormal{opt}$) based on SE- or WOT-SRSH, compared to literature reference values ($E^{\textnormal{lit.}}_\textnormal{opt}$). The optical band gap is dominated by the $\Gamma$ point for black phosphorus and by the $K$ point for MoS$_2$ and h-BN. Numbers in parentheses correspond to the position of the first absorption peak upon inclusion of spin-orbit coupling. Numbers in squared parentheses correspond to the position of the peaks upon inclusion of ZPR values (0.26 eV for bulk, 0.38 eV for 1L). \\
    \textsuperscript{A}Ref.\,\onlinecite{Tran2014}, value from GW-BSE.
    \textsuperscript{B}Ref.\,\onlinecite{Haastrup2018, Gjerding2021}, values from GW-BSE. 
    \textsuperscript{C}Ref. \onlinecite{Komsa2012}, experimental value.
    \textsuperscript{D}Ref.\,\onlinecite{Zhu2015}, experimental value. 
    \textsuperscript{E}Ref. \onlinecite{Tarrio1989}, experimental value. 
    \textsuperscript{F}Ref. \onlinecite{Louie2022}, value from GW-BSE.
    \textsuperscript{G}Ref. \onlinecite{Li2017_2}, experimental value.
    \textsuperscript{H}Ref. \onlinecite{Roman2021}, experimental value.
    }\label{t2}
\end{table*}

The above-explained WOT procedure is demonstrated in Fig.\ \ref{fig:method-IP-vdW}(b), which shows $\Delta I (\gamma)$ curves for various values of $\alpha$, for both bulk (3D $\epsilon_\infty$) and monolayer ($\epsilon_\infty = 1)$. 
The Figure shows that even over the relatively limited range of $\alpha$ and $\gamma$ values depicted, the bulk band gap can change by $\sim$1 eV and the monolayer gap can change by as much as $\sim$6 eV. This establishes the expected need for parameter tuning. 
Enforcing the IP ansatz, the spread of predicted band gap values is reduced to an insignificant $\sim$0.1 eV for the bulk, but in the monolayer the spread in band gap values is reduced to a smaller but not negligible $\sim$0.6 eV, which also affects computed optical spectra. 
Moreover, there are still many $\alpha$-$\gamma$ pairs for bulk or monolayer that satisfy the IP ansatz, and 
further unequivocal choice of $\alpha$ and $\gamma$ is needed. Therefore we introduce an additional optimization step, where we plot  curves in the $\alpha$-$\gamma$ plane that satisfy the IP ansatz of Eq.\ (\ref{eq:deltaI}), i.e. $\Delta I = 0$, for {\it both} bulk and monolayer. The curves, shown in Fig.\ \ref{fig:method-IP-vdW}(c), feature a {\it unique} crossing point, denoted by ($\alpha^*, \gamma^*$), which specifies optimal values that are both non-empirical {\it and} transferable between monolayer and bulk, thereby combining the advantages of WOT-SRSH for 3D solids with SE-SRSH for 2D materials, respectively.

\begin{figure*}
\centering
    \includegraphics[width=0.8\linewidth]{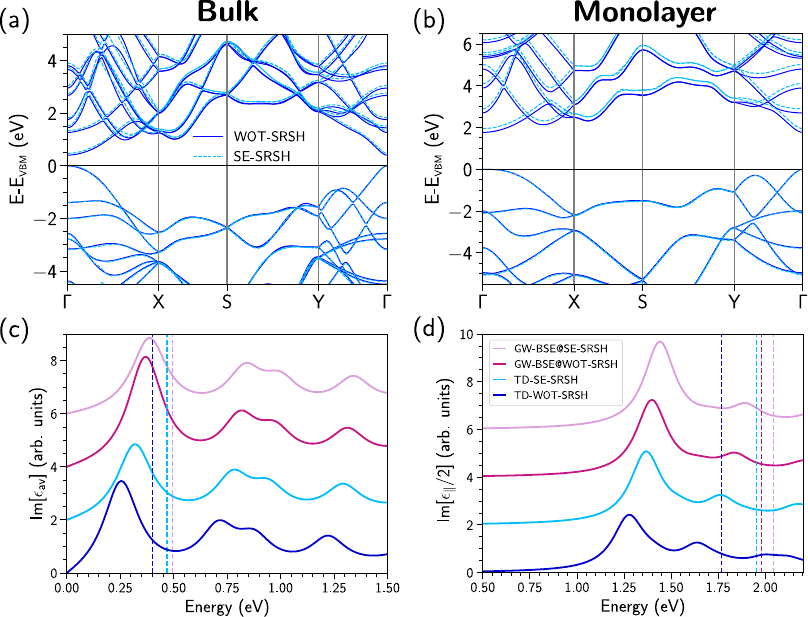}      
    \caption{ (a), (b) Bandstructures of bulk and monolayer black phosphorus, calculated using SE- and WOT-SRSH  functionals. (c), (d) Optical absorption spectra for bulk and monolayer black phosphorus, obtained from TD-SE-SRSH and TD-WOT-SRSH, as well as from ``single-shot'' GW-BSE using SE- and WOT-SRSH as starting points. Dashed vertical lines represent the fundamental band gap. SE-SRSH data are taken from Ref.\ \cite{Camarasa-Gomez2023}. 
    }  
\label{fig:BP_wot} 
\end{figure*}

\begin{figure*}
\centering
    \includegraphics[width=0.8\linewidth]{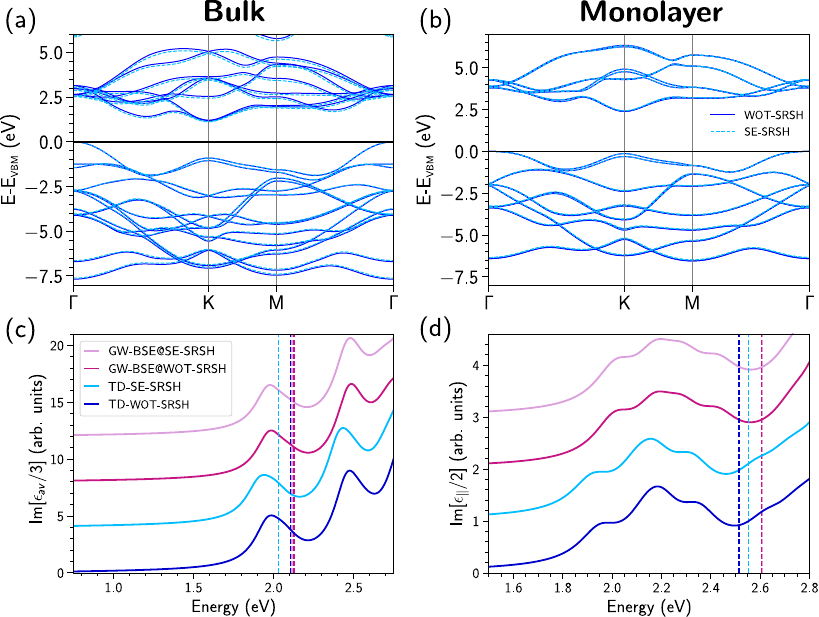}      
    \caption{ (a), (b) Bandstructures of bulk and monolayer MoS$_2$, calculated using SE- and WOT-SRSH  functionals \cite{note1}. (c), (d) Optical absorption spectra for bulk and monolayer MoS$_2$, obtained from TD-SE-SRSH and TD-WOT-SRSH, as well as from ``single-shot'' GW-BSE using SE- and WOT-SRSH as starting points. Dashed vertical lines represent the fundamental band gap. SE-SRSH data are taken from Ref.\ \cite{Camarasa-Gomez2023}. All monolayer calculations include spin-orbit coupling. 
    }
   \label{fig:MoS2_wot} 
\end{figure*}

We examine the accuracy of the proposed approach by applying it to bP, MoS$_2$, and h-BN, in both monolayer and bulk form - see Section II and III of the SI for computational details. First, we examine bP and  MoS$_2$, for which SE-SRSH has already produced accurate band structures and optical absorption spectra, albeit with empirical parameters \cite{Ramasubramaniam2019, Camarasa-Gomez2023}. We choose MoS$_2$ because it has been characterized extensively both experimentally and theoretically \cite{Wang2012, Geim2013, Liu2016, Novoselov2016, Ajayan2016, Geim2004, Geim2005, Radisavljevic2011, Lopez-Sanchez2013, Jariwala2014, Andrei2021}. We choose bP because its narrow ($\sim$0.3-0.6 eV), strain-sensitive \cite{Ling2015, Li2018, Xu2019, Cheng2020} bulk band gap provides a stringent performance test. For MoS$_2$, the generalized-gradient approximation of Perdew, Burke, and Ernzerhof (PBE)\cite{Perdew1996} has already proven to be a useful starting point for GW and for determining SE-SRSH parameters \cite{Ramasubramaniam2019}. It is therefore used here as well to calculate the dielectric constant in the initial step of the WOT-SRSH procedure. For bP, owing to the narrow bulk band gap, PBE is known to erroneously predict a metallic ground state \cite{Kotani2002, Schilfgaarde2006, Fuchs2007} and in this case we use the short-range hybrid functional of Heyd, Scuseria, and Ernzerhof (HSE){\cite{HSE, HSEerratum} instead of PBE \cite{Camarasa-Gomez2023}.

Using the above-described procedure, we obtain optimal $(\alpha^\ast,\gamma^\ast)$ pairs of (0.152, 0.027\AA$^{-1}$)  and (0.126, 0.030\AA$^{-1}$) for bP and MoS$_2$, respectively (see Section II of the SI for detailed parameters). These values differ (by 12-16\% for $\alpha^\ast$ and 24-26\% for $\gamma^\ast$)} from those obtained with the semiempirical procedure  (see SI),\cite{Ramasubramaniam2019, Camarasa-Gomez2023} which is expected given that the IP ansatz is satisfied in one case (WOT-SRSH) whereas target band gaps are fitted in the other (SE-SRSH). The band structures obtained from each method are compared in Figs.\ \ref{fig:BP_wot}(a-b) and \ref{fig:MoS2_wot}(a-b) for bP and MoS$_2$, respectively. Remarkably, despite their different parameters, WOT-SRSH and SE-SRSH produce bandstructures that are in strikingly good agreement. As the SE-SRSH band structures were already shown to be in excellent agreement with GW calculations \cite{Ramasubramaniam2019, Camarasa-Gomez2023}, so too are our WOT-SRSH results. Quantitatively, corresponding fundamental band gaps, also compared to literature values, are given in Table \ref{t1}. Once again, the SE- and WOT-SRSH values agree very well with each other (maximum deviation of $\sim$0.18 eV, for bulk MoS$_2$). 
Even more importantly, Table \ref{t1} shows that our results compare very well with reference literature data, i.e., are within the range of reported band gaps. This establishes the predictive power of our approach for fundamental band gaps and band structures. 

The calculation of accurate optical absorption spectra presents an even more stringent test for the predictive capacity of the SRSH functionals, as a correct description of exciton (de-)localization is needed. Results obtained from time-dependent SRSH (TD-SE-SRSH/TD-WOT-SRSH) calculations for bulk and monolayer bP and MoS$_2$ are given in Figs.\ \ref{fig:BP_wot}(c-d) and \ref{fig:MoS2_wot}(c-d), respectively. Clearly, the spectral lineshapes obtained with either approach, for both bulk and monolayer are essentially the same, with very small shifts between the two spectra (maximal shift of $\sim$ 0.15 eV, for monolayer bP). Furthermore, optical gaps, reported in Table \ref{t2}, are in excellent agreement with literature values. The Table also shows that for the MoS$_2$ monolayer we can accurately resolve the A and B spin-orbit split exciton peaks.

 To examine further the accuracy of the SRSH functionals, we perform  full-frequency, single-shot, G$_0$W$_0$ (henceforth, ``GW'') calculations,\cite{Shishkin2006} starting from the SE-/WOT-SRSH wavefunctions and eigenenergies, followed by BSE calculations for optical absorption spectra. The results are also shown in Figs.\ \ref{fig:BP_wot} (c-d) and \ref{fig:MoS2_wot}(c-d), with fundamental (GW) and optical (GW-BSE) gaps reported in Tables \ref{t1} and \ref{t2}; the dashed vertical lines in the figures indicate the calculated direct GW quasiparticle gaps. Importantly, the line-shapes obtained from TDDFT and from GW-BSE are essentially identical and quantitative shifts between the spectra obtained from either  methods are within the typical accuracy of either calculation, i.e., 0.1-0.2 eV.  
 The fact that GW corrections to WOT-SRSH change the band gap for bulk by $\sim$0.2 eV at most, and usually less (as found previously for non-layered materials \cite{Gant2022,Ohad2023}), reflects the quantitative accuracy of the WOT-SRSH approach. Thus, WOT-SRSH is shown to be equally useful in itself or as an optimal starting point for {\it ab initio} many-body perturbation theory \cite{Gant2022,Ohad2023}.

\begin{figure*}
\centering
    \includegraphics[width=0.8\linewidth]{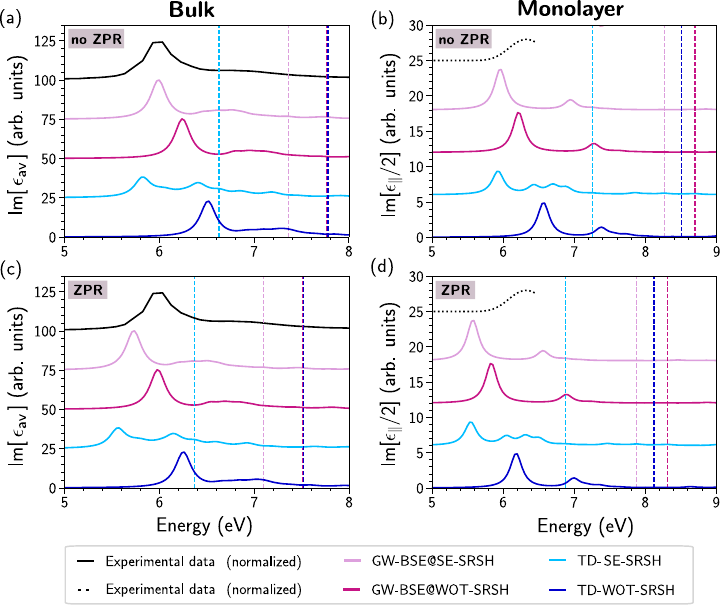}      
    \caption{Optical absorption spectra for (a, b) bulk and (c, d) monolayer h-BN. Experimental absorption spectra are extracted from Ref.\,\cite{Tarrio1989} for bulk h-BN and from Ref.\,\cite{Roman2021} for the monolayer. Dashed vertical lines represent the fundamental band gap. (a, c) and (b, d) show results without and with subtraction of ZPR corrections, respectively.  
    }\label{fig:hBN_wot} 
\end{figure*}

We now turn to h-BN, a wide-gap semiconductor. Particularly in monolayer form, it has proven challenging to predict accurately from first principles in general \cite{Hunt2020} and using SE-SRSH in particular \cite{Ramasubramaniam2019}. Recent work has reported that obtaining accurate quasiparticle gaps (and subsequent absorption spectra) for monolayer h-BN requires eigenvalue-self-consistent GW$_0$ calculations for a semilocal functional starting point.\cite{Louie2022} We therefore examine whether the WOT-SRSH method can overcome these challenges. 

Figure \ref{fig:hBN_wot}(a,b) displays optical absorption spectra for bulk and monolayer h-BN, obtained from TD-SE- and TD-WOT-SRSH calculations, as well as from GW-BSE calculations based on SE-/WOT-SRSH ground states as a starting point. Corresponding fundamental and optical band gaps are reported in Tables \ref{t1} and \ref{t2}, respectively. In agreement with Ref.\ \cite{Ramasubramaniam2019}, we find that the TD-SE-SRSH absorption spectra suffer from both qualitative and quantitative shortcomings, in the form of spurious satellite peaks and incorrect absorption onsets, respectively. These deficiencies are ameliorated by performing a GW calculation based on the SE-SRSH ground state and naively one could deduce that at least for the bulk (Figure \ref{fig:hBN_wot}), quantitative agreement with experiment is obtained, an issue we return to below. Importantly, spectra based on WOT-SRSH do not show a spurious line-shape even in the absence of GW-BSE corrections, thereby resolving a significant disadvantage of SE-SRSH while removing empiricism. And once again, spectral shifts upon application of GW-BSE from a WOT-SRSH starting point are small and the lineshapes are essentially the same.

Here, we note that electron-phonon interactions \cite{Allen1976, Allen1981} can lead to significant ZPR of electronic band gaps in materials with light elements, such as the B and N atoms found in h-BN \cite{Tutchton2018, Mishra2019, Hunt2020}. To account for this, we compute ZPR corrections to the fundamental gap of both bulk and monolayer h-BN using a finite difference approach \cite{Lloyd-williams2015,Monserrat2016,Monserrat2018} (see Sections I and II of the SI for details). 
We obtain ZPR shifts of 0.26 eV and 0.38 eV for the bulk and monolayer, respectively.
Computed optical spectra that are shifted by these ZPR values are given in Figure \ref{fig:hBN_wot}(c,d). For bulk h-BN, inclusion of the ZPR worsens the agreement between the GW-BSE@SE-SRSH absorption spectrum and experiment, indicating a partly fortuitous agreement in the absence of ZPR. For monolayer h-BN, GW-BSE@SE-SRSH calculations improves the spectral lineshape with respect to SE-SRSH, but does not fully address disagreement with experiment, which is already evident without ZPR corrections and only worsens upon inclusion of the latter.
In contrast, results based on the WOT-SRSH functional show improvement upon inclusion of the ZPR. 

The above discussion immediately clarifies the origins of the limitations of the SE-SRSH functional for h-BN \cite{Ramasubramaniam2019}, as well as why it is apparent only in this large gap material. The SE-SRSH was fit to a significantly underestimated fundamental gap (7.26 eV at the $K$ point) because GW with a local density approximation (LDA) starting point was used, whereas the WOT-SRSH band gap is substantially larger (8.13 eV at the $K$ point), consistent with self-consistent GW results \cite{Louie2022}. This underestimate by SE-SRSH and GW-BSE@PBE was partly offset by the neglect of ZPR, but errors remained. With WOT-SRSH, in contrast, the pertinent physical effects (single-particle excitations, exciton formation, phonon renormalization) are all accounted for systematically.  

In conclusion, we have presented a non-empirical and internally self-consistent approach to designing screened range-separated hybrid functionals that can deliver quantitatively accurate predictions of electronic and optical properties of layered van der Waals materials. These functionals already rival more expensive state-of-the-art methods such as GW-BSE at reduced computational cost and, moreover, if so desired can provide excellent starting points for higher-levels of theory. In particular, the straightforward resolution of the inaccuracies in quasiparticle and optical gaps for h-BN demonstrates the robustness of this approach and highlights its potential for modeling a broad range of 2D/layered semiconductors and insulators.

\begin{acknowledgments}
We thank Prof.\ Bartomeu Monserrat for very helpful input regarding the effects of ZPR and Dr.\ Daniel Hernang\'omez-P\'erez for useful discussions. This work was supported by the U.S.-Israel NSF–Binational Science Foundation Grant No. DMR-2015991, by the US Air Force through the grant AFOSR grant FA8655-20-1-7041, and by the Israel Science Foundation. M.C.-G. is grateful to the Azrieli Foundation for the award of an Azrieli International Postdoctoral Fellowship. This work used Frontera at TACC in part through allocation TG-DMR190070 from the Extreme Science and Engineering Discovery Environment (XSEDE) \cite{XSEDE}, which was supported by National Science Foundation grant number \#1548562 and the Advanced Cyberinfrastructure Coordination Ecosystem: Services \& Support (ACCESS) program, which is supported by National Science Foundation grants \#2138259, \#2138286, \#2138307, \#2137603, and \#2138296.
Additional computational resources were provided by the Weizmann Institute of Science at Chemfarm. L.K. acknowledges support from the Aryeh and Mintzi Katzman Professorial Chair, and the Helen and Martin Kimmel Award for Innovative Investigation. A.R. gratefully acknowledges support from the National Science Foundation (NSF-BSF 2150562).

\end{acknowledgments}

\bibliography{main}

\begin{thebibliography}{96}%
\makeatletter
\providecommand \@ifxundefined [1]{%
 \@ifx{#1\undefined}
}%
\providecommand \@ifnum [1]{%
 \ifnum #1\expandafter \@firstoftwo
 \else \expandafter \@secondoftwo
 \fi
}%
\providecommand \@ifx [1]{%
 \ifx #1\expandafter \@firstoftwo
 \else \expandafter \@secondoftwo
 \fi
}%
\providecommand \natexlab [1]{#1}%
\providecommand \enquote  [1]{``#1''}%
\providecommand \bibnamefont  [1]{#1}%
\providecommand \bibfnamefont [1]{#1}%
\providecommand \citenamefont [1]{#1}%
\providecommand \href@noop [0]{\@secondoftwo}%
\providecommand \href [0]{\begingroup \@sanitize@url \@href}%
\providecommand \@href[1]{\@@startlink{#1}\@@href}%
\providecommand \@@href[1]{\endgroup#1\@@endlink}%
\providecommand \@sanitize@url [0]{\catcode `\\12\catcode `\$12\catcode
  `\&12\catcode `\#12\catcode `\^12\catcode `\_12\catcode `\%12\relax}%
\providecommand \@@startlink[1]{}%
\providecommand \@@endlink[0]{}%
\providecommand \url  [0]{\begingroup\@sanitize@url \@url }%
\providecommand \@url [1]{\endgroup\@href {#1}{\urlprefix }}%
\providecommand \urlprefix  [0]{URL }%
\providecommand \Eprint [0]{\href }%
\providecommand \doibase [0]{https://doi.org/}%
\providecommand \selectlanguage [0]{\@gobble}%
\providecommand \bibinfo  [0]{\@secondoftwo}%
\providecommand \bibfield  [0]{\@secondoftwo}%
\providecommand \translation [1]{[#1]}%
\providecommand \BibitemOpen [0]{}%
\providecommand \bibitemStop [0]{}%
\providecommand \bibitemNoStop [0]{.\EOS\space}%
\providecommand \EOS [0]{\spacefactor3000\relax}%
\providecommand \BibitemShut  [1]{\csname bibitem#1\endcsname}%
\let\auto@bib@innerbib\@empty
\bibitem [{\citenamefont {Novoselov}\ \emph {et~al.}(2004)\citenamefont
  {Novoselov}, \citenamefont {Geim}, \citenamefont {Morozov}, \citenamefont
  {Jiang}, \citenamefont {Zhang}, \citenamefont {Dubonos}, \citenamefont
  {Grigorieva},\ and\ \citenamefont {Firsov}}]{Geim2004}%
  \BibitemOpen
  \bibfield  {author} {\bibinfo {author} {\bibfnamefont {K.~S.}\ \bibnamefont
  {Novoselov}}, \bibinfo {author} {\bibfnamefont {A.~K.}\ \bibnamefont {Geim}},
  \bibinfo {author} {\bibfnamefont {S.~V.}\ \bibnamefont {Morozov}}, \bibinfo
  {author} {\bibfnamefont {D.}~\bibnamefont {Jiang}}, \bibinfo {author}
  {\bibfnamefont {Y.}~\bibnamefont {Zhang}}, \bibinfo {author} {\bibfnamefont
  {S.~V.}\ \bibnamefont {Dubonos}}, \bibinfo {author} {\bibfnamefont {I.~V.}\
  \bibnamefont {Grigorieva}},\ and\ \bibinfo {author} {\bibfnamefont {A.~A.}\
  \bibnamefont {Firsov}},\ }\bibfield  {title} {\bibinfo {title} {Electric
  field effect in atomically thin carbon films},\ }\href
  {https://doi.org/10.1126/science.1102896} {\bibfield  {journal} {\bibinfo
  {journal} {Science}\ }\textbf {\bibinfo {volume} {306}},\ \bibinfo {pages}
  {666} (\bibinfo {year} {2004})}\BibitemShut {NoStop}%
\bibitem [{\citenamefont {Novoselov}\ \emph {et~al.}(2005)\citenamefont
  {Novoselov}, \citenamefont {Jiang}, \citenamefont {Schedin}, \citenamefont
  {Booth}, \citenamefont {Khotkevich}, \citenamefont {Morozov},\ and\
  \citenamefont {Geim}}]{Geim2005}%
  \BibitemOpen
  \bibfield  {author} {\bibinfo {author} {\bibfnamefont {K.~S.}\ \bibnamefont
  {Novoselov}}, \bibinfo {author} {\bibfnamefont {D.}~\bibnamefont {Jiang}},
  \bibinfo {author} {\bibfnamefont {F.}~\bibnamefont {Schedin}}, \bibinfo
  {author} {\bibfnamefont {T.~J.}\ \bibnamefont {Booth}}, \bibinfo {author}
  {\bibfnamefont {V.~V.}\ \bibnamefont {Khotkevich}}, \bibinfo {author}
  {\bibfnamefont {S.~V.}\ \bibnamefont {Morozov}},\ and\ \bibinfo {author}
  {\bibfnamefont {A.~K.}\ \bibnamefont {Geim}},\ }\bibfield  {title} {\bibinfo
  {title} {Two-dimensional atomic crystals},\ }\href
  {https://doi.org/10.1073/pnas.0502848102} {\bibfield  {journal} {\bibinfo
  {journal} {PNAS}\ }\textbf {\bibinfo {volume} {102}},\ \bibinfo {pages}
  {10451} (\bibinfo {year} {2005})}\BibitemShut {NoStop}%
\bibitem [{\citenamefont {Wang}\ \emph {et~al.}(2012)\citenamefont {Wang},
  \citenamefont {Kalantar-Zadeh}, \citenamefont {Kis}, \citenamefont
  {Coleman},\ and\ \citenamefont {Strano}}]{Wang2012}%
  \BibitemOpen
  \bibfield  {author} {\bibinfo {author} {\bibfnamefont {Q.~H.}\ \bibnamefont
  {Wang}}, \bibinfo {author} {\bibfnamefont {K.}~\bibnamefont
  {Kalantar-Zadeh}}, \bibinfo {author} {\bibfnamefont {A.}~\bibnamefont {Kis}},
  \bibinfo {author} {\bibfnamefont {J.~N.}\ \bibnamefont {Coleman}},\ and\
  \bibinfo {author} {\bibfnamefont {M.~S.}\ \bibnamefont {Strano}},\ }\bibfield
   {title} {\bibinfo {title} {Electronics and optoelectronics of
  two-dimensional transition metal dichalcogenides},\ }\href
  {https://doi.org/10.1038/nnano.2012.193} {\bibfield  {journal} {\bibinfo
  {journal} {Nature Nanotech.}\ }\textbf {\bibinfo {volume} {7}},\ \bibinfo
  {pages} {699} (\bibinfo {year} {2012})}\BibitemShut {NoStop}%
\bibitem [{\citenamefont {Geim}\ and\ \citenamefont
  {Grigorieva}(2013)}]{Geim2013}%
  \BibitemOpen
  \bibfield  {author} {\bibinfo {author} {\bibfnamefont {A.~K.}\ \bibnamefont
  {Geim}}\ and\ \bibinfo {author} {\bibfnamefont {I.~V.}\ \bibnamefont
  {Grigorieva}},\ }\bibfield  {title} {\bibinfo {title} {Van der {W}aals
  heterostructures},\ }\href {https://doi.org/10.1038/nature12385} {\bibfield
  {journal} {\bibinfo  {journal} {Nature}\ }\textbf {\bibinfo {volume} {499}},\
  \bibinfo {pages} {419} (\bibinfo {year} {2013})}\BibitemShut {NoStop}%
\bibitem [{\citenamefont {Liu}\ \emph {et~al.}(2016)\citenamefont {Liu},
  \citenamefont {Weiss}, \citenamefont {Duan}, \citenamefont {Cheng},
  \citenamefont {Huang},\ and\ \citenamefont {Duan}}]{Liu2016}%
  \BibitemOpen
  \bibfield  {author} {\bibinfo {author} {\bibfnamefont {Y.}~\bibnamefont
  {Liu}}, \bibinfo {author} {\bibfnamefont {N.~O.}\ \bibnamefont {Weiss}},
  \bibinfo {author} {\bibfnamefont {X.}~\bibnamefont {Duan}}, \bibinfo {author}
  {\bibfnamefont {H.-C.}\ \bibnamefont {Cheng}}, \bibinfo {author}
  {\bibfnamefont {Y.}~\bibnamefont {Huang}},\ and\ \bibinfo {author}
  {\bibfnamefont {X.}~\bibnamefont {Duan}},\ }\bibfield  {title} {\bibinfo
  {title} {Van der {W}aals heterostructures and devices},\ }\href
  {https://doi.org/10.1038/natrevmats.2016.42} {\bibfield  {journal} {\bibinfo
  {journal} {Nat. Rev. Mater.}\ }\textbf {\bibinfo {volume} {1}},\ \bibinfo
  {pages} {16042} (\bibinfo {year} {2016})}\BibitemShut {NoStop}%
\bibitem [{\citenamefont {Novoselov}\ \emph {et~al.}(2016)\citenamefont
  {Novoselov}, \citenamefont {Mishchenko}, \citenamefont {Carvalho},\ and\
  \citenamefont {Neto}}]{Novoselov2016}%
  \BibitemOpen
  \bibfield  {author} {\bibinfo {author} {\bibfnamefont {K.~S.}\ \bibnamefont
  {Novoselov}}, \bibinfo {author} {\bibfnamefont {A.}~\bibnamefont
  {Mishchenko}}, \bibinfo {author} {\bibfnamefont {A.}~\bibnamefont
  {Carvalho}},\ and\ \bibinfo {author} {\bibfnamefont {A.~H.~C.}\ \bibnamefont
  {Neto}},\ }\bibfield  {title} {\bibinfo {title} {{2D} materials and van der
  {W}aals heterostructures},\ }\href {https://doi.org/10.1126/science.aac9439}
  {\bibfield  {journal} {\bibinfo  {journal} {Science}\ }\textbf {\bibinfo
  {volume} {353}},\ \bibinfo {pages} {aac9439} (\bibinfo {year}
  {2016})}\BibitemShut {NoStop}%
\bibitem [{\citenamefont {Ajayan}\ \emph {et~al.}(2016)\citenamefont {Ajayan},
  \citenamefont {Kim},\ and\ \citenamefont {Banerjee}}]{Ajayan2016}%
  \BibitemOpen
  \bibfield  {author} {\bibinfo {author} {\bibfnamefont {P.~M.}\ \bibnamefont
  {Ajayan}}, \bibinfo {author} {\bibfnamefont {P.}~\bibnamefont {Kim}},\ and\
  \bibinfo {author} {\bibfnamefont {K.}~\bibnamefont {Banerjee}},\ }\bibfield
  {title} {\bibinfo {title} {Two-dimensional van der {W}aals materials},\
  }\href {https://doi.org/10.1063/PT.3.3297} {\bibfield  {journal} {\bibinfo
  {journal} {Phys. Today}\ }\textbf {\bibinfo {volume} {69}},\ \bibinfo {pages}
  {38} (\bibinfo {year} {2016})}\BibitemShut {NoStop}%
\bibitem [{\citenamefont {Bernardi}\ \emph {et~al.}(2017)\citenamefont
  {Bernardi}, \citenamefont {Ataca}, \citenamefont {Palummo},\ and\
  \citenamefont {Grossman}}]{Bernardi2017}%
  \BibitemOpen
  \bibfield  {author} {\bibinfo {author} {\bibfnamefont {M.}~\bibnamefont
  {Bernardi}}, \bibinfo {author} {\bibfnamefont {C.}~\bibnamefont {Ataca}},
  \bibinfo {author} {\bibfnamefont {M.}~\bibnamefont {Palummo}},\ and\ \bibinfo
  {author} {\bibfnamefont {J.~C.}\ \bibnamefont {Grossman}},\ }\bibfield
  {title} {\bibinfo {title} {Optical and electronic properties of
  two-dimensional layered materials},\ }\href
  {https://doi.org/doi:10.1515/nanoph-2015-0030} {\bibfield  {journal}
  {\bibinfo  {journal} {Nanophotonics}\ }\textbf {\bibinfo {volume} {6}},\
  \bibinfo {pages} {479} (\bibinfo {year} {2017})}\BibitemShut {NoStop}%
\bibitem [{\citenamefont {Yadav}\ \emph {et~al.}(2023)\citenamefont {Yadav},
  \citenamefont {Acosta}, \citenamefont {Dalpian},\ and\ \citenamefont
  {Malyi}}]{Yadav2023}%
  \BibitemOpen
  \bibfield  {author} {\bibinfo {author} {\bibfnamefont {A.}~\bibnamefont
  {Yadav}}, \bibinfo {author} {\bibfnamefont {C.~M.}\ \bibnamefont {Acosta}},
  \bibinfo {author} {\bibfnamefont {G.~M.}\ \bibnamefont {Dalpian}},\ and\
  \bibinfo {author} {\bibfnamefont {O.~I.}\ \bibnamefont {Malyi}},\ }\bibfield
  {title} {\bibinfo {title} {First-principles investigations of {2D} materials:
  Challenges and best practices},\ }\href
  {https://www.sciencedirect.com/science/article/pii/S2590238523002370}
  {\bibfield  {journal} {\bibinfo  {journal} {Matter}\ }\textbf {\bibinfo
  {volume} {6}},\ \bibinfo {pages} {2711} (\bibinfo {year} {2023})}\BibitemShut
  {NoStop}%
\bibitem [{\citenamefont {Hedin}(1965)}]{Hedin1965}%
  \BibitemOpen
  \bibfield  {author} {\bibinfo {author} {\bibfnamefont {L.}~\bibnamefont
  {Hedin}},\ }\bibfield  {title} {\bibinfo {title} {New method for calculating
  the one-particle {G}reen's function with application to the electron-gas
  problem},\ }\href {https://doi.org/10.1103/PhysRev.139.A796} {\bibfield
  {journal} {\bibinfo  {journal} {Phys. Rev.}\ }\textbf {\bibinfo {volume}
  {139}},\ \bibinfo {pages} {A796} (\bibinfo {year} {1965})}\BibitemShut
  {NoStop}%
\bibitem [{\citenamefont {Onida}\ \emph {et~al.}(2002)\citenamefont {Onida},
  \citenamefont {Reining},\ and\ \citenamefont {Rubio}}]{Onida2002}%
  \BibitemOpen
  \bibfield  {author} {\bibinfo {author} {\bibfnamefont {G.}~\bibnamefont
  {Onida}}, \bibinfo {author} {\bibfnamefont {L.}~\bibnamefont {Reining}},\
  and\ \bibinfo {author} {\bibfnamefont {A.}~\bibnamefont {Rubio}},\ }\bibfield
   {title} {\bibinfo {title} {{Electronic excitations: density-functional
  versus many-body Green's-function approaches}},\ }\href
  {https://doi.org/10.1103/RevModPhys.74.601} {\bibfield  {journal} {\bibinfo
  {journal} {Rev. Mod. Phys.}\ }\textbf {\bibinfo {volume} {74}},\ \bibinfo
  {pages} {601} (\bibinfo {year} {2002})}\BibitemShut {NoStop}%
\bibitem [{\citenamefont {Hybertsen}\ and\ \citenamefont
  {Louie}(1986)}]{Hybertsen1986}%
  \BibitemOpen
  \bibfield  {author} {\bibinfo {author} {\bibfnamefont {M.~S.}\ \bibnamefont
  {Hybertsen}}\ and\ \bibinfo {author} {\bibfnamefont {S.~G.}\ \bibnamefont
  {Louie}},\ }\bibfield  {title} {\bibinfo {title} {{Electron correlation in
  semiconductors and insulators: Band gaps and quasiparticle energies}},\
  }\href {https://doi.org/10.1103/PhysRevB.34.5390} {\bibfield  {journal}
  {\bibinfo  {journal} {Phys. Rev. B}\ }\textbf {\bibinfo {volume} {34}},\
  \bibinfo {pages} {5390} (\bibinfo {year} {1986})}\BibitemShut {NoStop}%
\bibitem [{\citenamefont {Rohlfing}\ and\ \citenamefont
  {Louie}(1998)}]{Rohlfing1998}%
  \BibitemOpen
  \bibfield  {author} {\bibinfo {author} {\bibfnamefont {M.}~\bibnamefont
  {Rohlfing}}\ and\ \bibinfo {author} {\bibfnamefont {S.~G.}\ \bibnamefont
  {Louie}},\ }\bibfield  {title} {\bibinfo {title} {Electron-hole excitations
  in semiconductors and insulators},\ }\href
  {https://doi.org/10.1103/PhysRevLett.81.2312} {\bibfield  {journal} {\bibinfo
   {journal} {Phys. Rev. Lett.}\ }\textbf {\bibinfo {volume} {81}},\ \bibinfo
  {pages} {2312} (\bibinfo {year} {1998})}\BibitemShut {NoStop}%
\bibitem [{\citenamefont {Albrecht}\ \emph {et~al.}(1998)\citenamefont
  {Albrecht}, \citenamefont {Reining}, \citenamefont {Del~Sole},\ and\
  \citenamefont {Onida}}]{Onida1998}%
  \BibitemOpen
  \bibfield  {author} {\bibinfo {author} {\bibfnamefont {S.}~\bibnamefont
  {Albrecht}}, \bibinfo {author} {\bibfnamefont {L.}~\bibnamefont {Reining}},
  \bibinfo {author} {\bibfnamefont {R.}~\bibnamefont {Del~Sole}},\ and\
  \bibinfo {author} {\bibfnamefont {G.}~\bibnamefont {Onida}},\ }\bibfield
  {title} {\bibinfo {title} {Ab initio calculation of excitonic effects in the
  optical spectra of semiconductors},\ }\href
  {https://doi.org/10.1103/PhysRevLett.80.4510} {\bibfield  {journal} {\bibinfo
   {journal} {Phys. Rev. Lett.}\ }\textbf {\bibinfo {volume} {80}},\ \bibinfo
  {pages} {4510} (\bibinfo {year} {1998})}\BibitemShut {NoStop}%
\bibitem [{\citenamefont {Parr}\ and\ \citenamefont {Yang}(1989)}]{Parr1989}%
  \BibitemOpen
  \bibfield  {author} {\bibinfo {author} {\bibfnamefont {R.~G.}\ \bibnamefont
  {Parr}}\ and\ \bibinfo {author} {\bibfnamefont {W.}~\bibnamefont {Yang}},\
  }\bibinfo {title} {{Density Functional Theory of Atoms and Molecules}}\
  (\bibinfo  {publisher} {Oxford University Press, Oxford},\ \bibinfo {year}
  {1989})\BibitemShut {NoStop}%
\bibitem [{\citenamefont {Dreizler}\ and\ \citenamefont
  {Gross}(1990)}]{Dreizler1990}%
  \BibitemOpen
  \bibfield  {author} {\bibinfo {author} {\bibfnamefont {M.}~\bibnamefont
  {Dreizler}}\ and\ \bibinfo {author} {\bibfnamefont {E.~K.~U.}\ \bibnamefont
  {Gross}},\ }\bibinfo {title} {{Density Functional Theory: An Approach to the
  Quantum Many-Body Problem}}\ (\bibinfo  {publisher} {Springer, Berlin},\
  \bibinfo {year} {1990})\BibitemShut {NoStop}%
\bibitem [{\citenamefont {Ma}\ and\ \citenamefont {Wang}(2016)}]{Ma2016}%
  \BibitemOpen
  \bibfield  {author} {\bibinfo {author} {\bibfnamefont {J.}~\bibnamefont
  {Ma}}\ and\ \bibinfo {author} {\bibfnamefont {L.-W.}\ \bibnamefont {Wang}},\
  }\bibfield  {title} {\bibinfo {title} {Using {W}annier functions to improve
  solid band gap predictions in density functional theory},\ }\href
  {https://doi.org/10.1038/srep24924} {\bibfield  {journal} {\bibinfo
  {journal} {Sci. Rep.}\ }\textbf {\bibinfo {volume} {6}},\ \bibinfo {pages}
  {24924} (\bibinfo {year} {2016})}\BibitemShut {NoStop}%
\bibitem [{\citenamefont {Weng}\ \emph {et~al.}(2017)\citenamefont {Weng},
  \citenamefont {Li}, \citenamefont {Ma}, \citenamefont {Zheng}, \citenamefont
  {Pan},\ and\ \citenamefont {Wang}}]{Weng2017}%
  \BibitemOpen
  \bibfield  {author} {\bibinfo {author} {\bibfnamefont {M.}~\bibnamefont
  {Weng}}, \bibinfo {author} {\bibfnamefont {S.}~\bibnamefont {Li}}, \bibinfo
  {author} {\bibfnamefont {J.}~\bibnamefont {Ma}}, \bibinfo {author}
  {\bibfnamefont {J.}~\bibnamefont {Zheng}}, \bibinfo {author} {\bibfnamefont
  {F.}~\bibnamefont {Pan}},\ and\ \bibinfo {author} {\bibfnamefont {L.-W.}\
  \bibnamefont {Wang}},\ }\bibfield  {title} {\bibinfo {title} {Wannier
  {K}oopman method calculations of the band gaps of alkali halides},\
  }\href@noop {} {\bibfield  {journal} {\bibinfo  {journal} {Appl. Phys.
  Lett.}\ }\textbf {\bibinfo {volume} {111}},\ \bibinfo {pages} {054101}
  (\bibinfo {year} {2017})}\BibitemShut {NoStop}%
\bibitem [{\citenamefont {Nguyen}\ \emph {et~al.}(2018)\citenamefont {Nguyen},
  \citenamefont {Colonna}, \citenamefont {Ferretti},\ and\ \citenamefont
  {Marzari}}]{Nguyen2018}%
  \BibitemOpen
  \bibfield  {author} {\bibinfo {author} {\bibfnamefont {N.~L.}\ \bibnamefont
  {Nguyen}}, \bibinfo {author} {\bibfnamefont {N.}~\bibnamefont {Colonna}},
  \bibinfo {author} {\bibfnamefont {A.}~\bibnamefont {Ferretti}},\ and\
  \bibinfo {author} {\bibfnamefont {N.}~\bibnamefont {Marzari}},\ }\bibfield
  {title} {\bibinfo {title} {Koopmans-compliant spectral functionals for
  extended systems},\ }\href@noop {} {\bibfield  {journal} {\bibinfo  {journal}
  {Phys. Rev. X}\ }\textbf {\bibinfo {volume} {8}},\ \bibinfo {pages} {021051}
  (\bibinfo {year} {2018})}\BibitemShut {NoStop}%
\bibitem [{\citenamefont {Colonna}\ \emph {et~al.}(2019)\citenamefont
  {Colonna}, \citenamefont {Nguyen}, \citenamefont {Ferretti},\ and\
  \citenamefont {Marzari}}]{Colonna2019}%
  \BibitemOpen
  \bibfield  {author} {\bibinfo {author} {\bibfnamefont {N.}~\bibnamefont
  {Colonna}}, \bibinfo {author} {\bibfnamefont {N.~L.}\ \bibnamefont {Nguyen}},
  \bibinfo {author} {\bibfnamefont {A.}~\bibnamefont {Ferretti}},\ and\
  \bibinfo {author} {\bibfnamefont {N.}~\bibnamefont {Marzari}},\ }\bibfield
  {title} {\bibinfo {title} {Koopmans-compliant functionals and potentials and
  their application to the {GW100} test set},\ }\href
  {https://doi.org/10.1021/acs.jctc.8b00976} {\bibfield  {journal} {\bibinfo
  {journal} {J. Chem. Theory Comput.}\ }\textbf {\bibinfo {volume} {15}},\
  \bibinfo {pages} {1905} (\bibinfo {year} {2019})}\BibitemShut {NoStop}%
\bibitem [{\citenamefont {Colonna}\ \emph {et~al.}(2022)\citenamefont
  {Colonna}, \citenamefont {De~Gennaro}, \citenamefont {Linscott},\ and\
  \citenamefont {Marzari}}]{Colonna2022}%
  \BibitemOpen
  \bibfield  {author} {\bibinfo {author} {\bibfnamefont {N.}~\bibnamefont
  {Colonna}}, \bibinfo {author} {\bibfnamefont {R.}~\bibnamefont {De~Gennaro}},
  \bibinfo {author} {\bibfnamefont {E.}~\bibnamefont {Linscott}},\ and\
  \bibinfo {author} {\bibfnamefont {N.}~\bibnamefont {Marzari}},\ }\bibfield
  {title} {\bibinfo {title} {Koopmans spectral functionals in periodic boundary
  conditions},\ }\href {https://doi.org/10.1021/acs.jctc.2c00161} {\bibfield
  {journal} {\bibinfo  {journal} {J. Chem. Theory Comput.}\ }\textbf {\bibinfo
  {volume} {18}},\ \bibinfo {pages} {5435} (\bibinfo {year}
  {2022})}\BibitemShut {NoStop}%
\bibitem [{\citenamefont {Linscott}\ \emph {et~al.}(2023)\citenamefont
  {Linscott}, \citenamefont {Colonna}, \citenamefont {De~Gennaro},
  \citenamefont {Nguyen}, \citenamefont {Borghi}, \citenamefont {Ferretti},
  \citenamefont {Dabo},\ and\ \citenamefont {Marzari}}]{Linscott2023}%
  \BibitemOpen
  \bibfield  {author} {\bibinfo {author} {\bibfnamefont {E.~B.}\ \bibnamefont
  {Linscott}}, \bibinfo {author} {\bibfnamefont {N.}~\bibnamefont {Colonna}},
  \bibinfo {author} {\bibfnamefont {R.}~\bibnamefont {De~Gennaro}}, \bibinfo
  {author} {\bibfnamefont {N.~L.}\ \bibnamefont {Nguyen}}, \bibinfo {author}
  {\bibfnamefont {G.}~\bibnamefont {Borghi}}, \bibinfo {author} {\bibfnamefont
  {A.}~\bibnamefont {Ferretti}}, \bibinfo {author} {\bibfnamefont
  {I.}~\bibnamefont {Dabo}},\ and\ \bibinfo {author} {\bibfnamefont
  {N.}~\bibnamefont {Marzari}},\ }\bibfield  {title} {\bibinfo {title}
  {koopmans: An open-source package for accurately and efficiently predicting
  spectral properties with koopmans functionals},\ }\href
  {https://doi.org/10.1021/acs.jctc.3c00652} {\bibfield  {journal} {\bibinfo
  {journal} {J. Chem. Theory Comput.}\ }\textbf {\bibinfo {volume} {19}},\
  \bibinfo {pages} {7097} (\bibinfo {year} {2023})}\BibitemShut {NoStop}%
\bibitem [{\citenamefont {Li}\ \emph {et~al.}(2018)\citenamefont {Li},
  \citenamefont {Sun}, \citenamefont {Shahi}, \citenamefont {Gao},
  \citenamefont {MacDonald}, \citenamefont {Uwatoko}, \citenamefont {Xiang},
  \citenamefont {Goodenough}, \citenamefont {Cheng},\ and\ \citenamefont
  {Zhou}}]{Li2018}%
  \BibitemOpen
  \bibfield  {author} {\bibinfo {author} {\bibfnamefont {X.}~\bibnamefont
  {Li}}, \bibinfo {author} {\bibfnamefont {J.}~\bibnamefont {Sun}}, \bibinfo
  {author} {\bibfnamefont {P.}~\bibnamefont {Shahi}}, \bibinfo {author}
  {\bibfnamefont {M.}~\bibnamefont {Gao}}, \bibinfo {author} {\bibfnamefont
  {A.~H.}\ \bibnamefont {MacDonald}}, \bibinfo {author} {\bibfnamefont
  {Y.}~\bibnamefont {Uwatoko}}, \bibinfo {author} {\bibfnamefont
  {T.}~\bibnamefont {Xiang}}, \bibinfo {author} {\bibfnamefont {J.~B.}\
  \bibnamefont {Goodenough}}, \bibinfo {author} {\bibfnamefont
  {J.}~\bibnamefont {Cheng}},\ and\ \bibinfo {author} {\bibfnamefont
  {J.}~\bibnamefont {Zhou}},\ }\bibfield  {title} {\bibinfo {title}
  {Pressure-induced phase transitions and superconductivity in a black
  phosphorus single crystal},\ }\href {https://doi.org/10.1073/pnas.1810726115}
  {\bibfield  {journal} {\bibinfo  {journal} {PNAS}\ }\textbf {\bibinfo
  {volume} {115}},\ \bibinfo {pages} {9935} (\bibinfo {year}
  {2018})}\BibitemShut {NoStop}%
\bibitem [{\citenamefont {Mei}\ \emph {et~al.}(2020)\citenamefont {Mei},
  \citenamefont {Chen},\ and\ \citenamefont {Yang}}]{Mei2020}%
  \BibitemOpen
  \bibfield  {author} {\bibinfo {author} {\bibfnamefont {Y.}~\bibnamefont
  {Mei}}, \bibinfo {author} {\bibfnamefont {Z.}~\bibnamefont {Chen}},\ and\
  \bibinfo {author} {\bibfnamefont {W.}~\bibnamefont {Yang}},\ }\bibfield
  {title} {\bibinfo {title} {Self-consistent calculation of the localized
  orbital scaling correction for correct electron densities and energy-level
  alignments in density functional theory},\ }\href
  {https://doi.org/10.1021/acs.jpclett.0c03133} {\bibfield  {journal} {\bibinfo
   {journal} {J. Phys. Chem. Lett.}\ }\textbf {\bibinfo {volume} {11}},\
  \bibinfo {pages} {10269} (\bibinfo {year} {2020})}\BibitemShut {NoStop}%
\bibitem [{\citenamefont {Mahler}\ \emph {et~al.}(2022)\citenamefont {Mahler},
  \citenamefont {Williams}, \citenamefont {Su},\ and\ \citenamefont
  {Yang}}]{Mahler2022}%
  \BibitemOpen
  \bibfield  {author} {\bibinfo {author} {\bibfnamefont {A.}~\bibnamefont
  {Mahler}}, \bibinfo {author} {\bibfnamefont {J.}~\bibnamefont {Williams}},
  \bibinfo {author} {\bibfnamefont {N.~Q.}\ \bibnamefont {Su}},\ and\ \bibinfo
  {author} {\bibfnamefont {W.}~\bibnamefont {Yang}},\ }\bibfield  {title}
  {\bibinfo {title} {Localized orbital scaling correction for periodic
  systems},\ }\href@noop {} {\bibfield  {journal} {\bibinfo  {journal} {Phys.
  Rev. B}\ }\textbf {\bibinfo {volume} {106}},\ \bibinfo {pages} {035147}
  (\bibinfo {year} {2022})}\BibitemShut {NoStop}%
\bibitem [{\citenamefont {Shimazaki}\ and\ \citenamefont
  {Asai}(2009)}]{Shimazaki2009}%
  \BibitemOpen
  \bibfield  {author} {\bibinfo {author} {\bibfnamefont {T.}~\bibnamefont
  {Shimazaki}}\ and\ \bibinfo {author} {\bibfnamefont {Y.}~\bibnamefont
  {Asai}},\ }\bibfield  {title} {\bibinfo {title} {{First principles band
  structure calculations based on self-consistent screened Hartree–Fock
  exchange potential}},\ }\href {https://doi.org/10.1063/1.3119259} {\bibfield
  {journal} {\bibinfo  {journal} {J. Chem. Phys.}\ }\textbf {\bibinfo {volume}
  {130}},\ \bibinfo {pages} {164702} (\bibinfo {year} {2009})}\BibitemShut
  {NoStop}%
\bibitem [{\citenamefont {Skone}\ \emph {et~al.}(2014)\citenamefont {Skone},
  \citenamefont {Govoni},\ and\ \citenamefont {Galli}}]{Skone2014}%
  \BibitemOpen
  \bibfield  {author} {\bibinfo {author} {\bibfnamefont {J.~H.}\ \bibnamefont
  {Skone}}, \bibinfo {author} {\bibfnamefont {M.}~\bibnamefont {Govoni}},\ and\
  \bibinfo {author} {\bibfnamefont {G.}~\bibnamefont {Galli}},\ }\bibfield
  {title} {\bibinfo {title} {Self-consistent hybrid functional for condensed
  systems},\ }\href {https://doi.org/10.1103/PhysRevB.89.195112} {\bibfield
  {journal} {\bibinfo  {journal} {Phys. Rev. B}\ }\textbf {\bibinfo {volume}
  {89}},\ \bibinfo {pages} {195112} (\bibinfo {year} {2014})}\BibitemShut
  {NoStop}%
\bibitem [{\citenamefont {Skone}\ \emph {et~al.}(2016)\citenamefont {Skone},
  \citenamefont {Govoni},\ and\ \citenamefont {Galli}}]{Skone2016}%
  \BibitemOpen
  \bibfield  {author} {\bibinfo {author} {\bibfnamefont {J.~H.}\ \bibnamefont
  {Skone}}, \bibinfo {author} {\bibfnamefont {M.}~\bibnamefont {Govoni}},\ and\
  \bibinfo {author} {\bibfnamefont {G.}~\bibnamefont {Galli}},\ }\bibfield
  {title} {\bibinfo {title} {Nonempirical range-separated hybrid functionals
  for solids and molecules},\ }\href
  {https://doi.org/10.1103/PhysRevB.93.235106} {\bibfield  {journal} {\bibinfo
  {journal} {Phys. Rev. B}\ }\textbf {\bibinfo {volume} {93}},\ \bibinfo
  {pages} {235106} (\bibinfo {year} {2016})}\BibitemShut {NoStop}%
\bibitem [{\citenamefont {Chen}\ \emph {et~al.}(2018)\citenamefont {Chen},
  \citenamefont {Miceli}, \citenamefont {Rignanese},\ and\ \citenamefont
  {Pasquarello}}]{Chen2018}%
  \BibitemOpen
  \bibfield  {author} {\bibinfo {author} {\bibfnamefont {W.}~\bibnamefont
  {Chen}}, \bibinfo {author} {\bibfnamefont {G.}~\bibnamefont {Miceli}},
  \bibinfo {author} {\bibfnamefont {G.-M.}\ \bibnamefont {Rignanese}},\ and\
  \bibinfo {author} {\bibfnamefont {A.}~\bibnamefont {Pasquarello}},\
  }\bibfield  {title} {\bibinfo {title} {Nonempirical dielectric-dependent
  hybrid functional with range separation for semiconductors and insulators},\
  }\href {https://doi.org/10.1103/PhysRevMaterials.2.073803} {\bibfield
  {journal} {\bibinfo  {journal} {Phys. Rev. Mater.}\ }\textbf {\bibinfo
  {volume} {2}},\ \bibinfo {pages} {073803} (\bibinfo {year}
  {2018})}\BibitemShut {NoStop}%
\bibitem [{\citenamefont {Sun}\ \emph {et~al.}(2020)\citenamefont {Sun},
  \citenamefont {Yang},\ and\ \citenamefont {Ullrich}}]{Sun2020}%
  \BibitemOpen
  \bibfield  {author} {\bibinfo {author} {\bibfnamefont {J.}~\bibnamefont
  {Sun}}, \bibinfo {author} {\bibfnamefont {J.}~\bibnamefont {Yang}},\ and\
  \bibinfo {author} {\bibfnamefont {C.~A.}\ \bibnamefont {Ullrich}},\
  }\bibfield  {title} {\bibinfo {title} {Low-cost alternatives to the
  {B}ethe-{S}alpeter equation: Towards simple hybrid functionals for excitonic
  effects in solids},\ }\href
  {https://doi.org/10.1103/PhysRevResearch.2.013091} {\bibfield  {journal}
  {\bibinfo  {journal} {Phys. Rev. Res.}\ }\textbf {\bibinfo {volume} {2}},\
  \bibinfo {pages} {013091} (\bibinfo {year} {2020})}\BibitemShut {NoStop}%
\bibitem [{\citenamefont {Zhan}\ \emph {et~al.}(2023)\citenamefont {Zhan},
  \citenamefont {Govoni},\ and\ \citenamefont {Galli}}]{Galli2023}%
  \BibitemOpen
  \bibfield  {author} {\bibinfo {author} {\bibfnamefont {J.}~\bibnamefont
  {Zhan}}, \bibinfo {author} {\bibfnamefont {M.}~\bibnamefont {Govoni}},\ and\
  \bibinfo {author} {\bibfnamefont {G.}~\bibnamefont {Galli}},\ }\bibfield
  {title} {\bibinfo {title} {Nonempirical range-separated hybrid functional
  with spatially dependent screened exchange},\ }\href
  {https://doi.org/10.1021/acs.jctc.3c00580} {\bibfield  {journal} {\bibinfo
  {journal} {J. Chem. Theory Comput.}\ }\textbf {\bibinfo {volume} {19}},\
  \bibinfo {pages} {5851} (\bibinfo {year} {2023})}\BibitemShut {NoStop}%
\bibitem [{\citenamefont {Refaely-Abramson}\ \emph {et~al.}(2013)\citenamefont
  {Refaely-Abramson}, \citenamefont {Sharifzadeh}, \citenamefont {Jain},
  \citenamefont {Baer}, \citenamefont {Neaton},\ and\ \citenamefont
  {Kronik}}]{Refaely2013}%
  \BibitemOpen
  \bibfield  {author} {\bibinfo {author} {\bibfnamefont {S.}~\bibnamefont
  {Refaely-Abramson}}, \bibinfo {author} {\bibfnamefont {S.}~\bibnamefont
  {Sharifzadeh}}, \bibinfo {author} {\bibfnamefont {M.}~\bibnamefont {Jain}},
  \bibinfo {author} {\bibfnamefont {R.}~\bibnamefont {Baer}}, \bibinfo {author}
  {\bibfnamefont {J.~B.}\ \bibnamefont {Neaton}},\ and\ \bibinfo {author}
  {\bibfnamefont {L.}~\bibnamefont {Kronik}},\ }\bibfield  {title} {\bibinfo
  {title} {{Gap renormalization of molecular crystals from density-functional
  theory}},\ }\href {https://doi.org/10.1103/PhysRevB.88.081204} {\bibfield
  {journal} {\bibinfo  {journal} {Phys. Rev. B}\ }\textbf {\bibinfo {volume}
  {88}},\ \bibinfo {pages} {081204} (\bibinfo {year} {2013})}\BibitemShut
  {NoStop}%
\bibitem [{\citenamefont {Refaely-Abramson}\ \emph {et~al.}(2015)\citenamefont
  {Refaely-Abramson}, \citenamefont {Jain}, \citenamefont {Sharifzadeh},
  \citenamefont {Neaton},\ and\ \citenamefont {Kronik}}]{Refaely2015}%
  \BibitemOpen
  \bibfield  {author} {\bibinfo {author} {\bibfnamefont {S.}~\bibnamefont
  {Refaely-Abramson}}, \bibinfo {author} {\bibfnamefont {M.}~\bibnamefont
  {Jain}}, \bibinfo {author} {\bibfnamefont {S.}~\bibnamefont {Sharifzadeh}},
  \bibinfo {author} {\bibfnamefont {J.~B.}\ \bibnamefont {Neaton}},\ and\
  \bibinfo {author} {\bibfnamefont {L.}~\bibnamefont {Kronik}},\ }\bibfield
  {title} {\bibinfo {title} {{Solid-state optical absorption from optimally
  tuned time-dependent range-separated hybrid density functional theory}},\
  }\href {https://doi.org/10.1103/PhysRevB.92.081204} {\bibfield  {journal}
  {\bibinfo  {journal} {Phys. Rev. B}\ }\textbf {\bibinfo {volume} {92}},\
  \bibinfo {pages} {081204} (\bibinfo {year} {2015})}\BibitemShut {NoStop}%
\bibitem [{\citenamefont {Miceli}\ \emph {et~al.}(2018)\citenamefont {Miceli},
  \citenamefont {Chen}, \citenamefont {Reshetnyak},\ and\ \citenamefont
  {Pasquarello}}]{Miceli2018}%
  \BibitemOpen
  \bibfield  {author} {\bibinfo {author} {\bibfnamefont {G.}~\bibnamefont
  {Miceli}}, \bibinfo {author} {\bibfnamefont {W.}~\bibnamefont {Chen}},
  \bibinfo {author} {\bibfnamefont {I.}~\bibnamefont {Reshetnyak}},\ and\
  \bibinfo {author} {\bibfnamefont {A.}~\bibnamefont {Pasquarello}},\
  }\bibfield  {title} {\bibinfo {title} {Nonempirical hybrid functionals for
  band gaps and polaronic distortions in solids},\ }\href
  {https://doi.org/10.1103/PhysRevB.97.121112} {\bibfield  {journal} {\bibinfo
  {journal} {Phys. Rev. B}\ }\textbf {\bibinfo {volume} {97}},\ \bibinfo
  {pages} {121112} (\bibinfo {year} {2018})}\BibitemShut {NoStop}%
\bibitem [{\citenamefont {Bischoff}\ \emph {et~al.}(2019)\citenamefont
  {Bischoff}, \citenamefont {Reshetnyak},\ and\ \citenamefont
  {Pasquarello}}]{Bischoff2019}%
  \BibitemOpen
  \bibfield  {author} {\bibinfo {author} {\bibfnamefont {T.}~\bibnamefont
  {Bischoff}}, \bibinfo {author} {\bibfnamefont {I.}~\bibnamefont
  {Reshetnyak}},\ and\ \bibinfo {author} {\bibfnamefont {A.}~\bibnamefont
  {Pasquarello}},\ }\bibfield  {title} {\bibinfo {title} {Adjustable potential
  probes for band-gap predictions of extended systems through nonempirical
  hybrid functionals},\ }\href@noop {} {\bibfield  {journal} {\bibinfo
  {journal} {Phys. Rev. B}\ }\textbf {\bibinfo {volume} {99}},\ \bibinfo
  {pages} {201114} (\bibinfo {year} {2019})}\BibitemShut {NoStop}%
\bibitem [{\citenamefont {Yang}\ \emph {et~al.}(2023)\citenamefont {Yang},
  \citenamefont {Falletta},\ and\ \citenamefont {Pasquarello}}]{Yang2023}%
  \BibitemOpen
  \bibfield  {author} {\bibinfo {author} {\bibfnamefont {J.}~\bibnamefont
  {Yang}}, \bibinfo {author} {\bibfnamefont {S.}~\bibnamefont {Falletta}},\
  and\ \bibinfo {author} {\bibfnamefont {A.}~\bibnamefont {Pasquarello}},\
  }\bibfield  {title} {\bibinfo {title} {Range-separated hybrid functionals for
  accurate prediction of band gaps of extended systems},\ }\href
  {https://doi.org/10.1038/s41524-023-01064-x} {\bibfield  {journal} {\bibinfo
  {journal} {{NPJ} Comp. Mater.}\ }\textbf {\bibinfo {volume} {9}},\ \bibinfo
  {pages} {108} (\bibinfo {year} {2023})}\BibitemShut {NoStop}%
\bibitem [{\citenamefont {Wing}\ \emph {et~al.}(2021)\citenamefont {Wing},
  \citenamefont {Ohad}, \citenamefont {Haber}, \citenamefont {Filip},
  \citenamefont {Gant}, \citenamefont {Neaton},\ and\ \citenamefont
  {Kronik}}]{Wing2021}%
  \BibitemOpen
  \bibfield  {author} {\bibinfo {author} {\bibfnamefont {D.}~\bibnamefont
  {Wing}}, \bibinfo {author} {\bibfnamefont {G.}~\bibnamefont {Ohad}}, \bibinfo
  {author} {\bibfnamefont {J.~B.}\ \bibnamefont {Haber}}, \bibinfo {author}
  {\bibfnamefont {M.~R.}\ \bibnamefont {Filip}}, \bibinfo {author}
  {\bibfnamefont {S.~E.}\ \bibnamefont {Gant}}, \bibinfo {author}
  {\bibfnamefont {J.~B.}\ \bibnamefont {Neaton}},\ and\ \bibinfo {author}
  {\bibfnamefont {L.}~\bibnamefont {Kronik}},\ }\bibfield  {title} {\bibinfo
  {title} {Band gaps of crystalline solids from {W}annier-localization based
  optimal tuning of a screened range-separated hybrid functional},\ }\href
  {https://doi.org/10.1073/pnas.2104556118} {\bibfield  {journal} {\bibinfo
  {journal} {PNAS}\ }\textbf {\bibinfo {volume} {118}},\ \bibinfo {pages}
  {e2104556118} (\bibinfo {year} {2021})}\BibitemShut {NoStop}%
\bibitem [{\citenamefont {Ohad}\ \emph {et~al.}(2022)\citenamefont {Ohad},
  \citenamefont {Wing}, \citenamefont {Gant}, \citenamefont {Cohen},
  \citenamefont {Haber}, \citenamefont {Sagredo}, \citenamefont {Filip},
  \citenamefont {Neaton},\ and\ \citenamefont {Kronik}}]{Ohad2022}%
  \BibitemOpen
  \bibfield  {author} {\bibinfo {author} {\bibfnamefont {G.}~\bibnamefont
  {Ohad}}, \bibinfo {author} {\bibfnamefont {D.}~\bibnamefont {Wing}}, \bibinfo
  {author} {\bibfnamefont {S.~E.}\ \bibnamefont {Gant}}, \bibinfo {author}
  {\bibfnamefont {A.~V.}\ \bibnamefont {Cohen}}, \bibinfo {author}
  {\bibfnamefont {J.~B.}\ \bibnamefont {Haber}}, \bibinfo {author}
  {\bibfnamefont {F.}~\bibnamefont {Sagredo}}, \bibinfo {author} {\bibfnamefont
  {M.~R.}\ \bibnamefont {Filip}}, \bibinfo {author} {\bibfnamefont {J.~B.}\
  \bibnamefont {Neaton}},\ and\ \bibinfo {author} {\bibfnamefont
  {L.}~\bibnamefont {Kronik}},\ }\bibfield  {title} {\bibinfo {title} {Band
  gaps of halide perovskites from a {W}annier-localized optimally tuned
  screened range-separated hybrid functional},\ }\href
  {https://doi.org/10.1103/PhysRevMaterials.6.104606} {\bibfield  {journal}
  {\bibinfo  {journal} {Phys. Rev. Materials}\ }\textbf {\bibinfo {volume}
  {6}},\ \bibinfo {pages} {104606} (\bibinfo {year} {2022})}\BibitemShut
  {NoStop}%
\bibitem [{\citenamefont {Gant}\ \emph {et~al.}(2022)\citenamefont {Gant},
  \citenamefont {Haber}, \citenamefont {Filip}, \citenamefont {Sagredo},
  \citenamefont {Wing}, \citenamefont {Ohad}, \citenamefont {Kronik},\ and\
  \citenamefont {Neaton}}]{Gant2022}%
  \BibitemOpen
  \bibfield  {author} {\bibinfo {author} {\bibfnamefont {S.~E.}\ \bibnamefont
  {Gant}}, \bibinfo {author} {\bibfnamefont {J.~B.}\ \bibnamefont {Haber}},
  \bibinfo {author} {\bibfnamefont {M.~R.}\ \bibnamefont {Filip}}, \bibinfo
  {author} {\bibfnamefont {F.}~\bibnamefont {Sagredo}}, \bibinfo {author}
  {\bibfnamefont {D.}~\bibnamefont {Wing}}, \bibinfo {author} {\bibfnamefont
  {G.}~\bibnamefont {Ohad}}, \bibinfo {author} {\bibfnamefont {L.}~\bibnamefont
  {Kronik}},\ and\ \bibinfo {author} {\bibfnamefont {J.~B.}\ \bibnamefont
  {Neaton}},\ }\bibfield  {title} {\bibinfo {title} {Optimally tuned starting
  point for single-shot {GW} calculations of solids},\ }\href
  {https://doi.org/10.1103/PhysRevMaterials.6.053802} {\bibfield  {journal}
  {\bibinfo  {journal} {Phys. Rev. Mater.}\ }\textbf {\bibinfo {volume} {6}},\
  \bibinfo {pages} {053802} (\bibinfo {year} {2022})}\BibitemShut {NoStop}%
\bibitem [{\citenamefont {Ohad}\ \emph {et~al.}(2023)\citenamefont {Ohad},
  \citenamefont {Gant}, \citenamefont {Wing}, \citenamefont {Haber},
  \citenamefont {Camarasa-G\'omez}, \citenamefont {Sagredo}, \citenamefont
  {Filip}, \citenamefont {Neaton},\ and\ \citenamefont {Kronik}}]{Ohad2023}%
  \BibitemOpen
  \bibfield  {author} {\bibinfo {author} {\bibfnamefont {G.}~\bibnamefont
  {Ohad}}, \bibinfo {author} {\bibfnamefont {S.~E.}\ \bibnamefont {Gant}},
  \bibinfo {author} {\bibfnamefont {D.}~\bibnamefont {Wing}}, \bibinfo {author}
  {\bibfnamefont {J.~B.}\ \bibnamefont {Haber}}, \bibinfo {author}
  {\bibfnamefont {M.}~\bibnamefont {Camarasa-G\'omez}}, \bibinfo {author}
  {\bibfnamefont {F.}~\bibnamefont {Sagredo}}, \bibinfo {author} {\bibfnamefont
  {M.~R.}\ \bibnamefont {Filip}}, \bibinfo {author} {\bibfnamefont {J.~B.}\
  \bibnamefont {Neaton}},\ and\ \bibinfo {author} {\bibfnamefont
  {L.}~\bibnamefont {Kronik}},\ }\bibfield  {title} {\bibinfo {title} {Optical
  absorption spectra of metal oxides from time-dependent density functional
  theory and many-body perturbation theory based on optimally-tuned hybrid
  functionals},\ }\href {https://doi.org/10.1103/PhysRevMaterials.7.123803}
  {\bibfield  {journal} {\bibinfo  {journal} {Phys. Rev. Mater.}\ }\textbf
  {\bibinfo {volume} {7}},\ \bibinfo {pages} {123803} (\bibinfo {year}
  {2023})}\BibitemShut {NoStop}%
\bibitem [{\citenamefont {Ohad}\ \emph {et~al.}(2024)\citenamefont {Ohad},
  \citenamefont {Hartstein}, \citenamefont {Gould}, \citenamefont {Neaton},\
  and\ \citenamefont {Kronik}}]{Ohad2024}%
  \BibitemOpen
  \bibfield  {author} {\bibinfo {author} {\bibfnamefont {G.}~\bibnamefont
  {Ohad}}, \bibinfo {author} {\bibfnamefont {M.}~\bibnamefont {Hartstein}},
  \bibinfo {author} {\bibfnamefont {T.}~\bibnamefont {Gould}}, \bibinfo
  {author} {\bibfnamefont {J.~B.}\ \bibnamefont {Neaton}},\ and\ \bibinfo
  {author} {\bibfnamefont {L.}~\bibnamefont {Kronik}},\ }\href@noop {}
  {\bibinfo {title} {Non-empirical prediction of the length-dependent
  ionization potential in molecular chains}} (\bibinfo {year} {2024}),\ \Eprint
  {https://arxiv.org/abs/2403.18518} {arXiv:2403.18518 [physics.chem-ph]}
  \BibitemShut {NoStop}%
\bibitem [{\citenamefont {Kronik}\ and\ \citenamefont
  {Neaton}(2016)}]{Kronik2016}%
  \BibitemOpen
  \bibfield  {author} {\bibinfo {author} {\bibfnamefont {L.}~\bibnamefont
  {Kronik}}\ and\ \bibinfo {author} {\bibfnamefont {J.~B.}\ \bibnamefont
  {Neaton}},\ }\bibfield  {title} {\bibinfo {title} {Excited-state properties
  of molecular solids from first principles},\ }\href
  {https://doi.org/10.1146/annurev-physchem-040214-121351} {\bibfield
  {journal} {\bibinfo  {journal} {Annu. Rev. Phys. Chem.}\ }\textbf {\bibinfo
  {volume} {67}},\ \bibinfo {pages} {587} (\bibinfo {year} {2016})}\BibitemShut
  {NoStop}%
\bibitem [{\citenamefont {Kronik}\ and\ \citenamefont
  {Kümmel}(2018)}]{Kronik2018}%
  \BibitemOpen
  \bibfield  {author} {\bibinfo {author} {\bibfnamefont {L.}~\bibnamefont
  {Kronik}}\ and\ \bibinfo {author} {\bibfnamefont {S.}~\bibnamefont
  {Kümmel}},\ }\bibfield  {title} {\bibinfo {title} {Dielectric screening
  meets optimally tuned density functionals},\ }\href
  {https://doi.org/10.1002/adma.201706560} {\bibfield  {journal} {\bibinfo
  {journal} {Adv. Mater.}\ }\textbf {\bibinfo {volume} {30}},\ \bibinfo {pages}
  {1706560} (\bibinfo {year} {2018})}\BibitemShut {NoStop}%
\bibitem [{\citenamefont {Ramasubramaniam}\ \emph {et~al.}(2019)\citenamefont
  {Ramasubramaniam}, \citenamefont {Wing},\ and\ \citenamefont
  {Kronik}}]{Ramasubramaniam2019}%
  \BibitemOpen
  \bibfield  {author} {\bibinfo {author} {\bibfnamefont {A.}~\bibnamefont
  {Ramasubramaniam}}, \bibinfo {author} {\bibfnamefont {D.}~\bibnamefont
  {Wing}},\ and\ \bibinfo {author} {\bibfnamefont {L.}~\bibnamefont {Kronik}},\
  }\bibfield  {title} {\bibinfo {title} {{Transferable screened range-separated
  hybrids for layered materials: The cases of ${\mathrm{MoS}}_{2}$ and h-BN}},\
  }\href {https://doi.org/10.1103/PhysRevMaterials.3.084007} {\bibfield
  {journal} {\bibinfo  {journal} {Phys. Rev. Materials}\ }\textbf {\bibinfo
  {volume} {3}},\ \bibinfo {pages} {084007} (\bibinfo {year}
  {2019})}\BibitemShut {NoStop}%
\bibitem [{\citenamefont {Camarasa-G\'omez}\ \emph {et~al.}(2023)\citenamefont
  {Camarasa-G\'omez}, \citenamefont {Ramasubramaniam}, \citenamefont {Neaton},\
  and\ \citenamefont {Kronik}}]{Camarasa-Gomez2023}%
  \BibitemOpen
  \bibfield  {author} {\bibinfo {author} {\bibfnamefont {M.}~\bibnamefont
  {Camarasa-G\'omez}}, \bibinfo {author} {\bibfnamefont {A.}~\bibnamefont
  {Ramasubramaniam}}, \bibinfo {author} {\bibfnamefont {J.~B.}\ \bibnamefont
  {Neaton}},\ and\ \bibinfo {author} {\bibfnamefont {L.}~\bibnamefont
  {Kronik}},\ }\bibfield  {title} {\bibinfo {title} {Transferable screened
  range-separated hybrid functionals for electronic and optical properties of
  van der waals materials},\ }\href
  {https://doi.org/10.1103/PhysRevMaterials.7.104001} {\bibfield  {journal}
  {\bibinfo  {journal} {Phys. Rev. Mater.}\ }\textbf {\bibinfo {volume} {7}},\
  \bibinfo {pages} {104001} (\bibinfo {year} {2023})}\BibitemShut {NoStop}%
\bibitem [{\citenamefont {Haastrup}\ \emph {et~al.}(2018)\citenamefont
  {Haastrup}, \citenamefont {Strange}, \citenamefont {Pandey}, \citenamefont
  {Deilmann}, \citenamefont {Schmidt}, \citenamefont {Hinsche}, \citenamefont
  {Gjerding}, \citenamefont {Torelli}, \citenamefont {Larsen}, \citenamefont
  {Riis-Jensen}, \citenamefont {Gath}, \citenamefont {Jacobsen}, \citenamefont
  {Mortensen}, \citenamefont {Olsen},\ and\ \citenamefont
  {Thygesen}}]{Haastrup2018}%
  \BibitemOpen
  \bibfield  {author} {\bibinfo {author} {\bibfnamefont {S.}~\bibnamefont
  {Haastrup}}, \bibinfo {author} {\bibfnamefont {M.}~\bibnamefont {Strange}},
  \bibinfo {author} {\bibfnamefont {M.}~\bibnamefont {Pandey}}, \bibinfo
  {author} {\bibfnamefont {T.}~\bibnamefont {Deilmann}}, \bibinfo {author}
  {\bibfnamefont {P.~S.}\ \bibnamefont {Schmidt}}, \bibinfo {author}
  {\bibfnamefont {N.~F.}\ \bibnamefont {Hinsche}}, \bibinfo {author}
  {\bibfnamefont {M.~N.}\ \bibnamefont {Gjerding}}, \bibinfo {author}
  {\bibfnamefont {D.}~\bibnamefont {Torelli}}, \bibinfo {author} {\bibfnamefont
  {P.~M.}\ \bibnamefont {Larsen}}, \bibinfo {author} {\bibfnamefont {A.~C.}\
  \bibnamefont {Riis-Jensen}}, \bibinfo {author} {\bibfnamefont
  {J.}~\bibnamefont {Gath}}, \bibinfo {author} {\bibfnamefont {K.~W.}\
  \bibnamefont {Jacobsen}}, \bibinfo {author} {\bibfnamefont {J.~J.}\
  \bibnamefont {Mortensen}}, \bibinfo {author} {\bibfnamefont {T.}~\bibnamefont
  {Olsen}},\ and\ \bibinfo {author} {\bibfnamefont {K.~S.}\ \bibnamefont
  {Thygesen}},\ }\bibfield  {title} {\bibinfo {title} {The computational {2D}
  materials database: high-throughput modeling and discovery of atomically thin
  crystals},\ }\href {https://doi.org/10.1088/2053-1583/aacfc1} {\bibfield
  {journal} {\bibinfo  {journal} {{2D} Materials}\ }\textbf {\bibinfo {volume}
  {5}},\ \bibinfo {pages} {042002} (\bibinfo {year} {2018})}\BibitemShut
  {NoStop}%
\bibitem [{\citenamefont {Zhang}\ \emph {et~al.}(2019)\citenamefont {Zhang},
  \citenamefont {Chen},\ and\ \citenamefont {Zhou}}]{Zhang2019}%
  \BibitemOpen
  \bibfield  {author} {\bibinfo {author} {\bibfnamefont {X.}~\bibnamefont
  {Zhang}}, \bibinfo {author} {\bibfnamefont {A.}~\bibnamefont {Chen}},\ and\
  \bibinfo {author} {\bibfnamefont {Z.}~\bibnamefont {Zhou}},\ }\bibfield
  {title} {\bibinfo {title} {High-throughput computational screening of layered
  and two-dimensional materials},\ }\href
  {https://doi.org/https://doi.org/10.1002/wcms.1385} {\bibfield  {journal}
  {\bibinfo  {journal} {WIREs Comp. Mol. Sci.}\ }\textbf {\bibinfo {volume}
  {9}},\ \bibinfo {pages} {e1385} (\bibinfo {year} {2019})}\BibitemShut
  {NoStop}%
\bibitem [{\citenamefont {Leininger}\ \emph {et~al.}(1997)\citenamefont
  {Leininger}, \citenamefont {Stoll}, \citenamefont {Werner},\ and\
  \citenamefont {Savin}}]{Leininger1997}%
  \BibitemOpen
  \bibfield  {author} {\bibinfo {author} {\bibfnamefont {T.}~\bibnamefont
  {Leininger}}, \bibinfo {author} {\bibfnamefont {H.}~\bibnamefont {Stoll}},
  \bibinfo {author} {\bibfnamefont {H.-J.}\ \bibnamefont {Werner}},\ and\
  \bibinfo {author} {\bibfnamefont {A.}~\bibnamefont {Savin}},\ }\bibfield
  {title} {\bibinfo {title} {{Combining long-range configuration interaction
  with short-range density functionals}},\ }\href
  {https://doi.org/https://doi.org/10.1016/S0009-2614(97)00758-6} {\bibfield
  {journal} {\bibinfo  {journal} {Chem. Phys. Lett.}\ }\textbf {\bibinfo
  {volume} {275}},\ \bibinfo {pages} {151} (\bibinfo {year}
  {1997})}\BibitemShut {NoStop}%
\bibitem [{\citenamefont {Yanai}\ \emph {et~al.}(2004)\citenamefont {Yanai},
  \citenamefont {Tew},\ and\ \citenamefont {Handy}}]{Yanai2004}%
  \BibitemOpen
  \bibfield  {author} {\bibinfo {author} {\bibfnamefont {T.}~\bibnamefont
  {Yanai}}, \bibinfo {author} {\bibfnamefont {D.~P.}\ \bibnamefont {Tew}},\
  and\ \bibinfo {author} {\bibfnamefont {N.~C.}\ \bibnamefont {Handy}},\
  }\bibfield  {title} {\bibinfo {title} {{A new hybrid exchange–correlation
  functional using the Coulomb-attenuating method (CAM-B3LYP)}},\ }\href
  {https://doi.org/https://doi.org/10.1016/j.cplett.2004.06.011} {\bibfield
  {journal} {\bibinfo  {journal} {Chem. Phys. Lett.}\ }\textbf {\bibinfo
  {volume} {393}},\ \bibinfo {pages} {51 } (\bibinfo {year}
  {2004})}\BibitemShut {NoStop}%
\bibitem [{\citenamefont {Wing}\ \emph {et~al.}(2019)\citenamefont {Wing},
  \citenamefont {Haber}, \citenamefont {Noff}, \citenamefont {Barker},
  \citenamefont {Egger}, \citenamefont {Ramasubramaniam}, \citenamefont
  {Louie}, \citenamefont {Neaton},\ and\ \citenamefont {Kronik}}]{Wing2019}%
  \BibitemOpen
  \bibfield  {author} {\bibinfo {author} {\bibfnamefont {D.}~\bibnamefont
  {Wing}}, \bibinfo {author} {\bibfnamefont {J.~B.}\ \bibnamefont {Haber}},
  \bibinfo {author} {\bibfnamefont {R.}~\bibnamefont {Noff}}, \bibinfo {author}
  {\bibfnamefont {B.}~\bibnamefont {Barker}}, \bibinfo {author} {\bibfnamefont
  {D.~A.}\ \bibnamefont {Egger}}, \bibinfo {author} {\bibfnamefont
  {A.}~\bibnamefont {Ramasubramaniam}}, \bibinfo {author} {\bibfnamefont
  {S.~G.}\ \bibnamefont {Louie}}, \bibinfo {author} {\bibfnamefont {J.~B.}\
  \bibnamefont {Neaton}},\ and\ \bibinfo {author} {\bibfnamefont
  {L.}~\bibnamefont {Kronik}},\ }\bibfield  {title} {\bibinfo {title}
  {{Comparing time-dependent density functional theory with many-body
  perturbation theory for semiconductors: Screened range-separated hybrids and
  the {GW} plus Bethe-Salpeter approach}},\ }\href
  {https://doi.org/10.1103/PhysRevMaterials.3.064603} {\bibfield  {journal}
  {\bibinfo  {journal} {Phys. Rev. Materials}\ }\textbf {\bibinfo {volume}
  {3}},\ \bibinfo {pages} {064603} (\bibinfo {year} {2019})}\BibitemShut
  {NoStop}%
\bibitem [{\citenamefont {Seidl}\ \emph {et~al.}(1996)\citenamefont {Seidl},
  \citenamefont {G\"orling}, \citenamefont {Vogl}, \citenamefont {Majewski},\
  and\ \citenamefont {Levy}}]{Seidl1996}%
  \BibitemOpen
  \bibfield  {author} {\bibinfo {author} {\bibfnamefont {A.}~\bibnamefont
  {Seidl}}, \bibinfo {author} {\bibfnamefont {A.}~\bibnamefont {G\"orling}},
  \bibinfo {author} {\bibfnamefont {P.}~\bibnamefont {Vogl}}, \bibinfo {author}
  {\bibfnamefont {J.~A.}\ \bibnamefont {Majewski}},\ and\ \bibinfo {author}
  {\bibfnamefont {M.}~\bibnamefont {Levy}},\ }\bibfield  {title} {\bibinfo
  {title} {Generalized {Kohn-Sham} schemes and the band-gap problem},\ }\href
  {https://doi.org/10.1103/PhysRevB.53.3764} {\bibfield  {journal} {\bibinfo
  {journal} {Phys. Rev. B}\ }\textbf {\bibinfo {volume} {53}},\ \bibinfo
  {pages} {3764} (\bibinfo {year} {1996})}\BibitemShut {NoStop}%
\bibitem [{\citenamefont {K\"ummel}\ and\ \citenamefont
  {Kronik}(2008)}]{Kummel2008}%
  \BibitemOpen
  \bibfield  {author} {\bibinfo {author} {\bibfnamefont {S.}~\bibnamefont
  {K\"ummel}}\ and\ \bibinfo {author} {\bibfnamefont {L.}~\bibnamefont
  {Kronik}},\ }\bibfield  {title} {\bibinfo {title} {Orbital-dependent density
  functionals: Theory and applications},\ }\href
  {https://doi.org/10.1103/RevModPhys.80.3} {\bibfield  {journal} {\bibinfo
  {journal} {Rev. Mod. Phys.}\ }\textbf {\bibinfo {volume} {80}},\ \bibinfo
  {pages} {3} (\bibinfo {year} {2008})}\BibitemShut {NoStop}%
\bibitem [{\citenamefont {Kronik}\ \emph {et~al.}(2012)\citenamefont {Kronik},
  \citenamefont {Stein}, \citenamefont {Refaely-Abramson},\ and\ \citenamefont
  {Baer}}]{Kronik2012}%
  \BibitemOpen
  \bibfield  {author} {\bibinfo {author} {\bibfnamefont {L.}~\bibnamefont
  {Kronik}}, \bibinfo {author} {\bibfnamefont {T.}~\bibnamefont {Stein}},
  \bibinfo {author} {\bibfnamefont {S.}~\bibnamefont {Refaely-Abramson}},\ and\
  \bibinfo {author} {\bibfnamefont {R.}~\bibnamefont {Baer}},\ }\bibfield
  {title} {\bibinfo {title} {{Excitation Gaps of Finite-Sized Systems from
  Optimally Tuned Range-Separated Hybrid Functionals}},\ }\href
  {https://doi.org/10.1021/ct2009363} {\bibfield  {journal} {\bibinfo
  {journal} {J. Chem. Theory Comput.}\ }\textbf {\bibinfo {volume} {8}},\
  \bibinfo {pages} {1515} (\bibinfo {year} {2012})}\BibitemShut {NoStop}%
\bibitem [{\citenamefont {Perdew}\ \emph {et~al.}(2017)\citenamefont {Perdew},
  \citenamefont {Yang}, \citenamefont {Burke}, \citenamefont {Yang},
  \citenamefont {Gross}, \citenamefont {Scheffler}, \citenamefont {Scuseria},
  \citenamefont {Henderson}, \citenamefont {Zhang}, \citenamefont {Ruzsinszky},
  \citenamefont {Peng}, \citenamefont {Sun}, \citenamefont {Trushin},\ and\
  \citenamefont {G\"orling}}]{Perdew2017}%
  \BibitemOpen
  \bibfield  {author} {\bibinfo {author} {\bibfnamefont {J.~P.}\ \bibnamefont
  {Perdew}}, \bibinfo {author} {\bibfnamefont {W.}~\bibnamefont {Yang}},
  \bibinfo {author} {\bibfnamefont {K.}~\bibnamefont {Burke}}, \bibinfo
  {author} {\bibfnamefont {Z.}~\bibnamefont {Yang}}, \bibinfo {author}
  {\bibfnamefont {E.~K.~U.}\ \bibnamefont {Gross}}, \bibinfo {author}
  {\bibfnamefont {M.}~\bibnamefont {Scheffler}}, \bibinfo {author}
  {\bibfnamefont {G.~E.}\ \bibnamefont {Scuseria}}, \bibinfo {author}
  {\bibfnamefont {T.~M.}\ \bibnamefont {Henderson}}, \bibinfo {author}
  {\bibfnamefont {I.~Y.}\ \bibnamefont {Zhang}}, \bibinfo {author}
  {\bibfnamefont {A.}~\bibnamefont {Ruzsinszky}}, \bibinfo {author}
  {\bibfnamefont {H.}~\bibnamefont {Peng}}, \bibinfo {author} {\bibfnamefont
  {J.}~\bibnamefont {Sun}}, \bibinfo {author} {\bibfnamefont {E.}~\bibnamefont
  {Trushin}},\ and\ \bibinfo {author} {\bibfnamefont {A.}~\bibnamefont
  {G\"orling}},\ }\bibfield  {title} {\bibinfo {title} {Understanding band gaps
  of solids in generalized {Kohn–Sham} theory},\ }\href
  {https://www.pnas.org/doi/abs/10.1073/pnas.1621352114} {\bibfield  {journal}
  {\bibinfo  {journal} {PNAS}\ }\textbf {\bibinfo {volume} {114}},\ \bibinfo
  {pages} {2801} (\bibinfo {year} {2017})}\BibitemShut {NoStop}%
\bibitem [{\citenamefont {Baer}\ and\ \citenamefont {Kronik}(2019)}]{Baer2018}%
  \BibitemOpen
  \bibfield  {author} {\bibinfo {author} {\bibfnamefont {R.}~\bibnamefont
  {Baer}}\ and\ \bibinfo {author} {\bibfnamefont {L.}~\bibnamefont {Kronik}},\
  }\bibfield  {title} {\bibinfo {title} {Time-dependent generalized
  {Kohn–Sham} theory},\ }\href {https://doi.org/10.1140/epjb/e2018-90103-0}
  {\bibfield  {journal} {\bibinfo  {journal} {Eur. Phys. J. B}\ }\textbf
  {\bibinfo {volume} {91}},\ \bibinfo {pages} {170} (\bibinfo {year}
  {2019})}\BibitemShut {NoStop}%
\bibitem [{\citenamefont {Cudazzo}\ \emph {et~al.}(2011)\citenamefont
  {Cudazzo}, \citenamefont {Tokatly},\ and\ \citenamefont
  {Rubio}}]{Cudazzo2011}%
  \BibitemOpen
  \bibfield  {author} {\bibinfo {author} {\bibfnamefont {P.}~\bibnamefont
  {Cudazzo}}, \bibinfo {author} {\bibfnamefont {I.~V.}\ \bibnamefont
  {Tokatly}},\ and\ \bibinfo {author} {\bibfnamefont {A.}~\bibnamefont
  {Rubio}},\ }\bibfield  {title} {\bibinfo {title} {Dielectric screening in
  two-dimensional insulators: Implications for excitonic and impurity states in
  graphane},\ }\href {https://link.aps.org/doi/10.1103/PhysRevB.84.085406}
  {\bibfield  {journal} {\bibinfo  {journal} {Phys. Rev. B}\ }\textbf {\bibinfo
  {volume} {84}},\ \bibinfo {pages} {085406} (\bibinfo {year}
  {2011})}\BibitemShut {NoStop}%
\bibitem [{\citenamefont {Andersen}\ \emph {et~al.}(2015)\citenamefont
  {Andersen}, \citenamefont {Latini},\ and\ \citenamefont
  {Thygesen}}]{Andersen2015}%
  \BibitemOpen
  \bibfield  {author} {\bibinfo {author} {\bibfnamefont {K.}~\bibnamefont
  {Andersen}}, \bibinfo {author} {\bibfnamefont {S.}~\bibnamefont {Latini}},\
  and\ \bibinfo {author} {\bibfnamefont {K.~S.}\ \bibnamefont {Thygesen}},\
  }\bibfield  {title} {\bibinfo {title} {Dielectric genome of van der {W}aals
  heterostructures},\ }\href {https://doi.org/10.1021/acs.nanolett.5b01251}
  {\bibfield  {journal} {\bibinfo  {journal} {Nano Lett.}\ }\textbf {\bibinfo
  {volume} {15}},\ \bibinfo {pages} {4616} (\bibinfo {year}
  {2015})}\BibitemShut {NoStop}%
\bibitem [{\citenamefont {Qiu}\ \emph {et~al.}(2016)\citenamefont {Qiu},
  \citenamefont {da~Jornada},\ and\ \citenamefont {Louie}}]{Qiu2016}%
  \BibitemOpen
  \bibfield  {author} {\bibinfo {author} {\bibfnamefont {D.~Y.}\ \bibnamefont
  {Qiu}}, \bibinfo {author} {\bibfnamefont {F.~H.}\ \bibnamefont
  {da~Jornada}},\ and\ \bibinfo {author} {\bibfnamefont {S.~G.}\ \bibnamefont
  {Louie}},\ }\bibfield  {title} {\bibinfo {title} {Screening and many-body
  effects in two-dimensional crystals: Monolayer {MoS\textsubscript{2}}},\
  }\href {https://doi.org/10.1103/PhysRevB.93.235435} {\bibfield  {journal}
  {\bibinfo  {journal} {Phys. Rev. B}\ }\textbf {\bibinfo {volume} {93}},\
  \bibinfo {pages} {235435} (\bibinfo {year} {2016})}\BibitemShut {NoStop}%
\bibitem [{\citenamefont {Nunes}\ and\ \citenamefont
  {Gonze}(2001)}]{Nunes2001}%
  \BibitemOpen
  \bibfield  {author} {\bibinfo {author} {\bibfnamefont {R.~W.}\ \bibnamefont
  {Nunes}}\ and\ \bibinfo {author} {\bibfnamefont {X.}~\bibnamefont {Gonze}},\
  }\bibfield  {title} {\bibinfo {title} {Berry-phase treatment of the
  homogeneous electric field perturbation in insulators},\ }\href
  {https://doi.org/10.1103/PhysRevB.63.155107} {\bibfield  {journal} {\bibinfo
  {journal} {Phys. Rev. B}\ }\textbf {\bibinfo {volume} {63}},\ \bibinfo
  {pages} {155107} (\bibinfo {year} {2001})}\BibitemShut {NoStop}%
\bibitem [{\citenamefont {Souza}\ \emph {et~al.}(2002)\citenamefont {Souza},
  \citenamefont {\'I\~niguez},\ and\ \citenamefont {Vanderbilt}}]{Souza2002}%
  \BibitemOpen
  \bibfield  {author} {\bibinfo {author} {\bibfnamefont {I.}~\bibnamefont
  {Souza}}, \bibinfo {author} {\bibfnamefont {J.}~\bibnamefont {\'I\~niguez}},\
  and\ \bibinfo {author} {\bibfnamefont {D.}~\bibnamefont {Vanderbilt}},\
  }\bibfield  {title} {\bibinfo {title} {First-principles approach to
  insulators in finite electric fields},\ }\href
  {https://doi.org/10.1103/PhysRevLett.89.117602} {\bibfield  {journal}
  {\bibinfo  {journal} {Phys. Rev. Lett.}\ }\textbf {\bibinfo {volume} {89}},\
  \bibinfo {pages} {117602} (\bibinfo {year} {2002})}\BibitemShut {NoStop}%
\bibitem [{not()}]{note1}%
  \BibitemOpen
  \href@noop {} {}\bibinfo {note} {The smallest gap for the monolayer is
  indirect, between $\Gamma$ and K, rather than direct at K due to a biaxial
  tensile strain of $\sim$1.15\% that arises from using the experimental
  lattice parameter. This has no bearing on the overall conclusions of this
  paper.}\BibitemShut {Stop}%
\bibitem [{\citenamefont {Wang}\ \emph {et~al.}(2015)\citenamefont {Wang},
  \citenamefont {Kawazoe},\ and\ \citenamefont {Geng}}]{Wang2015}%
  \BibitemOpen
  \bibfield  {author} {\bibinfo {author} {\bibfnamefont {V.}~\bibnamefont
  {Wang}}, \bibinfo {author} {\bibfnamefont {Y.}~\bibnamefont {Kawazoe}},\ and\
  \bibinfo {author} {\bibfnamefont {W.~T.}\ \bibnamefont {Geng}},\ }\bibfield
  {title} {\bibinfo {title} {Native point defects in few-layer phosphorene},\
  }\href {https://doi.org/10.1103/PhysRevB.91.045433} {\bibfield  {journal}
  {\bibinfo  {journal} {Phys. Rev. B}\ }\textbf {\bibinfo {volume} {91}},\
  \bibinfo {pages} {045433} (\bibinfo {year} {2015})}\BibitemShut {NoStop}%
\bibitem [{\citenamefont {Tran}\ \emph {et~al.}(2014)\citenamefont {Tran},
  \citenamefont {Soklaski}, \citenamefont {Liang},\ and\ \citenamefont
  {Yang}}]{Tran2014}%
  \BibitemOpen
  \bibfield  {author} {\bibinfo {author} {\bibfnamefont {V.}~\bibnamefont
  {Tran}}, \bibinfo {author} {\bibfnamefont {R.}~\bibnamefont {Soklaski}},
  \bibinfo {author} {\bibfnamefont {Y.}~\bibnamefont {Liang}},\ and\ \bibinfo
  {author} {\bibfnamefont {L.}~\bibnamefont {Yang}},\ }\bibfield  {title}
  {\bibinfo {title} {Layer-controlled band gap and anisotropic excitons in
  few-layer black phosphorus},\ }\href
  {https://doi.org/10.1103/PhysRevB.89.235319} {\bibfield  {journal} {\bibinfo
  {journal} {Phys. Rev. B}\ }\textbf {\bibinfo {volume} {89}},\ \bibinfo
  {pages} {235319} (\bibinfo {year} {2014})}\BibitemShut {NoStop}%
\bibitem [{\citenamefont {Gjerding}\ \emph {et~al.}(2021)\citenamefont
  {Gjerding}, \citenamefont {Taghizadeh}, \citenamefont {Rasmussen},
  \citenamefont {Ali}, \citenamefont {Bertoldo}, \citenamefont {Deilmann},
  \citenamefont {Kn{\o}sgaard}, \citenamefont {Kruse}, \citenamefont {Larsen},
  \citenamefont {Manti}, \citenamefont {Pedersen}, \citenamefont {Petralanda},
  \citenamefont {Skovhus}, \citenamefont {Svendsen}, \citenamefont {Mortensen},
  \citenamefont {Olsen},\ and\ \citenamefont {Thygesen}}]{Gjerding2021}%
  \BibitemOpen
  \bibfield  {author} {\bibinfo {author} {\bibfnamefont {M.~N.}\ \bibnamefont
  {Gjerding}}, \bibinfo {author} {\bibfnamefont {A.}~\bibnamefont
  {Taghizadeh}}, \bibinfo {author} {\bibfnamefont {A.}~\bibnamefont
  {Rasmussen}}, \bibinfo {author} {\bibfnamefont {S.}~\bibnamefont {Ali}},
  \bibinfo {author} {\bibfnamefont {F.}~\bibnamefont {Bertoldo}}, \bibinfo
  {author} {\bibfnamefont {T.}~\bibnamefont {Deilmann}}, \bibinfo {author}
  {\bibfnamefont {N.~R.}\ \bibnamefont {Kn{\o}sgaard}}, \bibinfo {author}
  {\bibfnamefont {M.}~\bibnamefont {Kruse}}, \bibinfo {author} {\bibfnamefont
  {A.~H.}\ \bibnamefont {Larsen}}, \bibinfo {author} {\bibfnamefont
  {S.}~\bibnamefont {Manti}}, \bibinfo {author} {\bibfnamefont {T.~G.}\
  \bibnamefont {Pedersen}}, \bibinfo {author} {\bibfnamefont {U.}~\bibnamefont
  {Petralanda}}, \bibinfo {author} {\bibfnamefont {T.}~\bibnamefont {Skovhus}},
  \bibinfo {author} {\bibfnamefont {M.~K.}\ \bibnamefont {Svendsen}}, \bibinfo
  {author} {\bibfnamefont {J.~J.}\ \bibnamefont {Mortensen}}, \bibinfo {author}
  {\bibfnamefont {T.}~\bibnamefont {Olsen}},\ and\ \bibinfo {author}
  {\bibfnamefont {K.~S.}\ \bibnamefont {Thygesen}},\ }\bibfield  {title}
  {\bibinfo {title} {Recent progress of the computational {2D} materials
  database (c2db)},\ }\href {https://doi.org/10.1088/2053-1583/ac1059}
  {\bibfield  {journal} {\bibinfo  {journal} {{2D} Materials}\ }\textbf
  {\bibinfo {volume} {8}},\ \bibinfo {pages} {044002} (\bibinfo {year}
  {2021})}\BibitemShut {NoStop}%
\bibitem [{\citenamefont {Komsa}\ and\ \citenamefont
  {Krasheninnikov}(2012)}]{Komsa2012}%
  \BibitemOpen
  \bibfield  {author} {\bibinfo {author} {\bibfnamefont {H.-P.}\ \bibnamefont
  {Komsa}}\ and\ \bibinfo {author} {\bibfnamefont {A.~V.}\ \bibnamefont
  {Krasheninnikov}},\ }\bibfield  {title} {\bibinfo {title} {Effects of
  confinement and environment on the electronic structure and exciton binding
  energy of mos${}_{2}$ from first principles},\ }\href
  {https://doi.org/10.1103/PhysRevB.86.241201} {\bibfield  {journal} {\bibinfo
  {journal} {Phys. Rev. B}\ }\textbf {\bibinfo {volume} {86}},\ \bibinfo
  {pages} {241201} (\bibinfo {year} {2012})}\BibitemShut {NoStop}%
\bibitem [{\citenamefont {Graml}\ \emph {et~al.}(2024)\citenamefont {Graml},
  \citenamefont {Zollner}, \citenamefont {Hernang\'{o}mez-P\'{e}rez},
  \citenamefont {Faria~Junior},\ and\ \citenamefont {Wilhelm}}]{Wilhelm2023}%
  \BibitemOpen
  \bibfield  {author} {\bibinfo {author} {\bibfnamefont {M.}~\bibnamefont
  {Graml}}, \bibinfo {author} {\bibfnamefont {K.}~\bibnamefont {Zollner}},
  \bibinfo {author} {\bibfnamefont {D.}~\bibnamefont
  {Hernang\'{o}mez-P\'{e}rez}}, \bibinfo {author} {\bibfnamefont {P.~E.}\
  \bibnamefont {Faria~Junior}},\ and\ \bibinfo {author} {\bibfnamefont
  {J.}~\bibnamefont {Wilhelm}},\ }\bibfield  {title} {\bibinfo {title}
  {Low-scaling {GW} algorithm applied to twisted transition-metal
  dichalcogenide heterobilayers},\ }\href
  {https://doi.org/10.1021/acs.jctc.3c01230} {\bibfield  {journal} {\bibinfo
  {journal} {J. Chem. Theory Comp.}\ }\textbf {\bibinfo {volume} {20}},\
  \bibinfo {pages} {2202} (\bibinfo {year} {2024})}\BibitemShut {NoStop}%
\bibitem [{\citenamefont {Qiu}\ \emph {et~al.}(2013)\citenamefont {Qiu},
  \citenamefont {da~Jornada},\ and\ \citenamefont {Louie}}]{Qiu2013}%
  \BibitemOpen
  \bibfield  {author} {\bibinfo {author} {\bibfnamefont {D.~Y.}\ \bibnamefont
  {Qiu}}, \bibinfo {author} {\bibfnamefont {F.~H.}\ \bibnamefont
  {da~Jornada}},\ and\ \bibinfo {author} {\bibfnamefont {S.~G.}\ \bibnamefont
  {Louie}},\ }\bibfield  {title} {\bibinfo {title} {Optical spectrum of
  ${\mathrm{mos}}_{2}$: Many-body effects and diversity of exciton states},\
  }\href {https://doi.org/10.1103/PhysRevLett.111.216805} {\bibfield  {journal}
  {\bibinfo  {journal} {Phys. Rev. Lett.}\ }\textbf {\bibinfo {volume} {111}},\
  \bibinfo {pages} {216805} (\bibinfo {year} {2013})}\BibitemShut {NoStop}%
\bibitem [{\citenamefont {Kolos}\ and\ \citenamefont
  {Karlick{\'y}}(2019)}]{Kolos2019}%
  \BibitemOpen
  \bibfield  {author} {\bibinfo {author} {\bibfnamefont {M.}~\bibnamefont
  {Kolos}}\ and\ \bibinfo {author} {\bibfnamefont {F.}~\bibnamefont
  {Karlick{\'y}}},\ }\bibfield  {title} {\bibinfo {title} {Accurate many-body
  calculation of electronic and optical band gap of bulk hexagonal boron
  nitride},\ }\href {https://doi.org/10.1039/C8CP07328G} {\bibfield  {journal}
  {\bibinfo  {journal} {Phys. Chem. Chem. Phys.}\ }\textbf {\bibinfo {volume}
  {21}},\ \bibinfo {pages} {3999} (\bibinfo {year} {2019})}\BibitemShut
  {NoStop}%
\bibitem [{\citenamefont {Zhang}\ \emph {et~al.}(2022)\citenamefont {Zhang},
  \citenamefont {Ong}, \citenamefont {Ruan}, \citenamefont {Wu}, \citenamefont
  {Shi}, \citenamefont {Tang},\ and\ \citenamefont {Louie}}]{Louie2022}%
  \BibitemOpen
  \bibfield  {author} {\bibinfo {author} {\bibfnamefont {F.}~\bibnamefont
  {Zhang}}, \bibinfo {author} {\bibfnamefont {C.~S.}\ \bibnamefont {Ong}},
  \bibinfo {author} {\bibfnamefont {J.~W.}\ \bibnamefont {Ruan}}, \bibinfo
  {author} {\bibfnamefont {M.}~\bibnamefont {Wu}}, \bibinfo {author}
  {\bibfnamefont {X.~Q.}\ \bibnamefont {Shi}}, \bibinfo {author} {\bibfnamefont
  {Z.~K.}\ \bibnamefont {Tang}},\ and\ \bibinfo {author} {\bibfnamefont
  {S.~G.}\ \bibnamefont {Louie}},\ }\bibfield  {title} {\bibinfo {title}
  {Intervalley excitonic hybridization, optical selection rules, and imperfect
  circular dichroism in monolayer $h\text{\ensuremath{-}}\mathrm{BN}$},\ }\href
  {https://doi.org/10.1103/PhysRevLett.128.047402} {\bibfield  {journal}
  {\bibinfo  {journal} {Phys. Rev. Lett.}\ }\textbf {\bibinfo {volume} {128}},\
  \bibinfo {pages} {047402} (\bibinfo {year} {2022})}\BibitemShut {NoStop}%
\bibitem [{\citenamefont {Zhu}\ \emph {et~al.}(2015)\citenamefont {Zhu},
  \citenamefont {Chen},\ and\ \citenamefont {Cui}}]{Zhu2015}%
  \BibitemOpen
  \bibfield  {author} {\bibinfo {author} {\bibfnamefont {B.}~\bibnamefont
  {Zhu}}, \bibinfo {author} {\bibfnamefont {X.}~\bibnamefont {Chen}},\ and\
  \bibinfo {author} {\bibfnamefont {X.}~\bibnamefont {Cui}},\ }\bibfield
  {title} {\bibinfo {title} {Exciton binding energy of monolayer
  {WS\textsubscript{2}}},\ }\href {https://doi.org/10.1038/srep09218}
  {\bibfield  {journal} {\bibinfo  {journal} {Sci. Rep.}\ }\textbf {\bibinfo
  {volume} {5}},\ \bibinfo {pages} {9218} (\bibinfo {year} {2015})}\BibitemShut
  {NoStop}%
\bibitem [{\citenamefont {Tarrio}\ and\ \citenamefont
  {Schnatterly}(1989)}]{Tarrio1989}%
  \BibitemOpen
  \bibfield  {author} {\bibinfo {author} {\bibfnamefont {C.}~\bibnamefont
  {Tarrio}}\ and\ \bibinfo {author} {\bibfnamefont {S.~E.}\ \bibnamefont
  {Schnatterly}},\ }\bibfield  {title} {\bibinfo {title} {Interband
  transitions, plasmons, and dispersion in hexagonal boron nitride},\ }\href
  {https://doi.org/10.1103/PhysRevB.40.7852} {\bibfield  {journal} {\bibinfo
  {journal} {Phys. Rev. B}\ }\textbf {\bibinfo {volume} {40}},\ \bibinfo
  {pages} {7852} (\bibinfo {year} {1989})}\BibitemShut {NoStop}%
\bibitem [{\citenamefont {Li}\ \emph {et~al.}(2017)\citenamefont {Li},
  \citenamefont {Qiu}, \citenamefont {Liu}, \citenamefont {Yin},\ and\
  \citenamefont {Guo}}]{Li2017_2}%
  \BibitemOpen
  \bibfield  {author} {\bibinfo {author} {\bibfnamefont {X.}~\bibnamefont
  {Li}}, \bibinfo {author} {\bibfnamefont {H.}~\bibnamefont {Qiu}}, \bibinfo
  {author} {\bibfnamefont {X.}~\bibnamefont {Liu}}, \bibinfo {author}
  {\bibfnamefont {J.}~\bibnamefont {Yin}},\ and\ \bibinfo {author}
  {\bibfnamefont {W.}~\bibnamefont {Guo}},\ }\bibfield  {title} {\bibinfo
  {title} {Wettability of supported monolayer hexagonal boron nitride in air},\
  }\href {https://doi.org/https://doi.org/10.1002/adfm.201603181} {\bibfield
  {journal} {\bibinfo  {journal} {Adv. Funct. Mater.}\ }\textbf {\bibinfo
  {volume} {27}},\ \bibinfo {pages} {1603181} (\bibinfo {year}
  {2017})}\BibitemShut {NoStop}%
\bibitem [{\citenamefont {Rom{á}n Pe\~{n}a}\ \emph {et~al.}(2021)\citenamefont
  {Rom{á}n Pe\~{n}a}, \citenamefont {Costa~Costa}, \citenamefont {Zobelli},
  \citenamefont {Elias}, \citenamefont {Valvin}, \citenamefont {Cassabois},
  \citenamefont {Gil}, \citenamefont {Summerfield}, \citenamefont {Cheng},
  \citenamefont {Mellor}, \citenamefont {Beton}, \citenamefont {Novikov},\ and\
  \citenamefont {Zagonel}}]{Roman2021}%
  \BibitemOpen
  \bibfield  {author} {\bibinfo {author} {\bibfnamefont {R.~J.~P.}\
  \bibnamefont {Rom{á}n Pe\~{n}a}}, \bibinfo {author} {\bibfnamefont
  {F.~J.~R.}\ \bibnamefont {Costa~Costa}}, \bibinfo {author} {\bibfnamefont
  {A.}~\bibnamefont {Zobelli}}, \bibinfo {author} {\bibfnamefont
  {C.}~\bibnamefont {Elias}}, \bibinfo {author} {\bibfnamefont
  {P.}~\bibnamefont {Valvin}}, \bibinfo {author} {\bibfnamefont
  {G.}~\bibnamefont {Cassabois}}, \bibinfo {author} {\bibfnamefont
  {B.}~\bibnamefont {Gil}}, \bibinfo {author} {\bibfnamefont {A.}~\bibnamefont
  {Summerfield}}, \bibinfo {author} {\bibfnamefont {T.~S.}\ \bibnamefont
  {Cheng}}, \bibinfo {author} {\bibfnamefont {C.~J.}\ \bibnamefont {Mellor}},
  \bibinfo {author} {\bibfnamefont {P.~H.}\ \bibnamefont {Beton}}, \bibinfo
  {author} {\bibfnamefont {S.~V.}\ \bibnamefont {Novikov}},\ and\ \bibinfo
  {author} {\bibfnamefont {L.~Z.}\ \bibnamefont {Zagonel}},\ }\bibfield
  {title} {\bibinfo {title} {Band gap measurements of monolayer {h-BN} and
  insights into carbon-related point defects},\ }\href
  {https://doi.org/10.1088/2053-1583/ac0d9c} {\bibfield  {journal} {\bibinfo
  {journal} {2D Materials}\ }\textbf {\bibinfo {volume} {8}},\ \bibinfo {pages}
  {044001} (\bibinfo {year} {2021})}\BibitemShut {NoStop}%
\bibitem [{\citenamefont {Radisavljevic}\ \emph {et~al.}(2011)\citenamefont
  {Radisavljevic}, \citenamefont {Radenovic}, \citenamefont {Brivio},
  \citenamefont {Giacometti},\ and\ \citenamefont {Kis}}]{Radisavljevic2011}%
  \BibitemOpen
  \bibfield  {author} {\bibinfo {author} {\bibfnamefont {B.}~\bibnamefont
  {Radisavljevic}}, \bibinfo {author} {\bibfnamefont {A.}~\bibnamefont
  {Radenovic}}, \bibinfo {author} {\bibfnamefont {J.}~\bibnamefont {Brivio}},
  \bibinfo {author} {\bibfnamefont {V.}~\bibnamefont {Giacometti}},\ and\
  \bibinfo {author} {\bibfnamefont {A.}~\bibnamefont {Kis}},\ }\bibfield
  {title} {\bibinfo {title} {Single-layer {MoS$_2$} transistors},\ }\href
  {https://doi.org/10.1038/nnano.2010.279} {\bibfield  {journal} {\bibinfo
  {journal} {Nature Nanotechnology}\ }\textbf {\bibinfo {volume} {6}},\
  \bibinfo {pages} {147} (\bibinfo {year} {2011})}\BibitemShut {NoStop}%
\bibitem [{\citenamefont {Lopez-Sanchez}\ \emph {et~al.}(2013)\citenamefont
  {Lopez-Sanchez}, \citenamefont {Lembke}, \citenamefont {Kayci}, \citenamefont
  {Radenovic},\ and\ \citenamefont {Kis}}]{Lopez-Sanchez2013}%
  \BibitemOpen
  \bibfield  {author} {\bibinfo {author} {\bibfnamefont {O.}~\bibnamefont
  {Lopez-Sanchez}}, \bibinfo {author} {\bibfnamefont {D.}~\bibnamefont
  {Lembke}}, \bibinfo {author} {\bibfnamefont {M.}~\bibnamefont {Kayci}},
  \bibinfo {author} {\bibfnamefont {A.}~\bibnamefont {Radenovic}},\ and\
  \bibinfo {author} {\bibfnamefont {A.}~\bibnamefont {Kis}},\ }\bibfield
  {title} {\bibinfo {title} {Ultrasensitive photodetectors based on monolayer
  {MoS$_2$}},\ }\href {https://doi.org/10.1038/nnano.2013.100} {\bibfield
  {journal} {\bibinfo  {journal} {Nature Nanotechnology}\ }\textbf {\bibinfo
  {volume} {8}},\ \bibinfo {pages} {497} (\bibinfo {year} {2013})}\BibitemShut
  {NoStop}%
\bibitem [{\citenamefont {Jariwala}\ \emph {et~al.}(2014)\citenamefont
  {Jariwala}, \citenamefont {Sangwan}, \citenamefont {Lauhon}, \citenamefont
  {Marks},\ and\ \citenamefont {Hersam}}]{Jariwala2014}%
  \BibitemOpen
  \bibfield  {author} {\bibinfo {author} {\bibfnamefont {D.}~\bibnamefont
  {Jariwala}}, \bibinfo {author} {\bibfnamefont {V.~K.}\ \bibnamefont
  {Sangwan}}, \bibinfo {author} {\bibfnamefont {L.~J.}\ \bibnamefont {Lauhon}},
  \bibinfo {author} {\bibfnamefont {T.~J.}\ \bibnamefont {Marks}},\ and\
  \bibinfo {author} {\bibfnamefont {M.~C.}\ \bibnamefont {Hersam}},\ }\bibfield
   {title} {\bibinfo {title} {Emerging device applications for semiconducting
  two-dimensional transition metal dichalcogenides},\ }\href
  {https://doi.org/10.1021/nn500064s} {\bibfield  {journal} {\bibinfo
  {journal} {ACS Nano}\ }\textbf {\bibinfo {volume} {8}},\ \bibinfo {pages}
  {1102} (\bibinfo {year} {2014})}\BibitemShut {NoStop}%
\bibitem [{\citenamefont {Andrei}\ \emph {et~al.}(2021)\citenamefont {Andrei},
  \citenamefont {Efetov}, \citenamefont {Jarillo-Herrero}, \citenamefont
  {MacDonald}, \citenamefont {Mak}, \citenamefont {Senthil}, \citenamefont
  {Tutuc}, \citenamefont {Yazdani},\ and\ \citenamefont {Young}}]{Andrei2021}%
  \BibitemOpen
  \bibfield  {author} {\bibinfo {author} {\bibfnamefont {E.~Y.}\ \bibnamefont
  {Andrei}}, \bibinfo {author} {\bibfnamefont {D.~K.}\ \bibnamefont {Efetov}},
  \bibinfo {author} {\bibfnamefont {P.}~\bibnamefont {Jarillo-Herrero}},
  \bibinfo {author} {\bibfnamefont {A.~H.}\ \bibnamefont {MacDonald}}, \bibinfo
  {author} {\bibfnamefont {K.~F.}\ \bibnamefont {Mak}}, \bibinfo {author}
  {\bibfnamefont {T.}~\bibnamefont {Senthil}}, \bibinfo {author} {\bibfnamefont
  {E.}~\bibnamefont {Tutuc}}, \bibinfo {author} {\bibfnamefont
  {A.}~\bibnamefont {Yazdani}},\ and\ \bibinfo {author} {\bibfnamefont {A.~F.}\
  \bibnamefont {Young}},\ }\bibfield  {title} {\bibinfo {title} {The marvels of
  {M}oir{\'e} materials},\ }\href {https://doi.org/10.1038/s41578-021-00284-1}
  {\bibfield  {journal} {\bibinfo  {journal} {Nat. Rev. Mater.}\ }\textbf
  {\bibinfo {volume} {6}},\ \bibinfo {pages} {201} (\bibinfo {year}
  {2021})}\BibitemShut {NoStop}%
\bibitem [{\citenamefont {Ling}\ \emph {et~al.}(2015)\citenamefont {Ling},
  \citenamefont {Wang}, \citenamefont {Huang}, \citenamefont {Xia},\ and\
  \citenamefont {Dresselhaus}}]{Ling2015}%
  \BibitemOpen
  \bibfield  {author} {\bibinfo {author} {\bibfnamefont {X.}~\bibnamefont
  {Ling}}, \bibinfo {author} {\bibfnamefont {H.}~\bibnamefont {Wang}}, \bibinfo
  {author} {\bibfnamefont {S.}~\bibnamefont {Huang}}, \bibinfo {author}
  {\bibfnamefont {F.}~\bibnamefont {Xia}},\ and\ \bibinfo {author}
  {\bibfnamefont {M.~S.}\ \bibnamefont {Dresselhaus}},\ }\bibfield  {title}
  {\bibinfo {title} {The renaissance of black phosphorus},\ }\href
  {https://doi.org/10.1073/pnas.1416581112} {\bibfield  {journal} {\bibinfo
  {journal} {PNAS}\ }\textbf {\bibinfo {volume} {112}},\ \bibinfo {pages}
  {4523} (\bibinfo {year} {2015})}\BibitemShut {NoStop}%
\bibitem [{\citenamefont {Xu}\ \emph {et~al.}(2019)\citenamefont {Xu},
  \citenamefont {Shi}, \citenamefont {Shi}, \citenamefont {Zhang},\ and\
  \citenamefont {Zhang}}]{Xu2019}%
  \BibitemOpen
  \bibfield  {author} {\bibinfo {author} {\bibfnamefont {Y.}~\bibnamefont
  {Xu}}, \bibinfo {author} {\bibfnamefont {Z.}~\bibnamefont {Shi}}, \bibinfo
  {author} {\bibfnamefont {X.}~\bibnamefont {Shi}}, \bibinfo {author}
  {\bibfnamefont {K.}~\bibnamefont {Zhang}},\ and\ \bibinfo {author}
  {\bibfnamefont {H.}~\bibnamefont {Zhang}},\ }\bibfield  {title} {\bibinfo
  {title} {Recent progress in black phosphorus and black-phosphorus-analogue
  materials: properties, synthesis and applications},\ }\href
  {https://doi.org/10.1039/C9NR04348A} {\bibfield  {journal} {\bibinfo
  {journal} {Nanoscale}\ }\textbf {\bibinfo {volume} {11}},\ \bibinfo {pages}
  {14491} (\bibinfo {year} {2019})}\BibitemShut {NoStop}%
\bibitem [{\citenamefont {Cheng}\ \emph {et~al.}(2020)\citenamefont {Cheng},
  \citenamefont {Gao}, \citenamefont {Li}, \citenamefont {Mei}, \citenamefont
  {Wang}, \citenamefont {Wen}, \citenamefont {Huang}, \citenamefont {Li},
  \citenamefont {Zheng}, \citenamefont {Wang},\ and\ \citenamefont
  {Zhang}}]{Cheng2020}%
  \BibitemOpen
  \bibfield  {author} {\bibinfo {author} {\bibfnamefont {J.}~\bibnamefont
  {Cheng}}, \bibinfo {author} {\bibfnamefont {L.}~\bibnamefont {Gao}}, \bibinfo
  {author} {\bibfnamefont {T.}~\bibnamefont {Li}}, \bibinfo {author}
  {\bibfnamefont {S.}~\bibnamefont {Mei}}, \bibinfo {author} {\bibfnamefont
  {C.}~\bibnamefont {Wang}}, \bibinfo {author} {\bibfnamefont {B.}~\bibnamefont
  {Wen}}, \bibinfo {author} {\bibfnamefont {W.}~\bibnamefont {Huang}}, \bibinfo
  {author} {\bibfnamefont {C.}~\bibnamefont {Li}}, \bibinfo {author}
  {\bibfnamefont {G.}~\bibnamefont {Zheng}}, \bibinfo {author} {\bibfnamefont
  {H.}~\bibnamefont {Wang}},\ and\ \bibinfo {author} {\bibfnamefont
  {H.}~\bibnamefont {Zhang}},\ }\bibfield  {title} {\bibinfo {title}
  {Two-dimensional black phosphorus nanomaterials: Emerging advances in
  electrochemical energy storage science},\ }\href
  {https://doi.org/10.1007/s40820-020-00510-5} {\bibfield  {journal} {\bibinfo
  {journal} {Nanomicro Lett.}\ }\textbf {\bibinfo {volume} {12}},\ \bibinfo
  {pages} {179} (\bibinfo {year} {2020})}\BibitemShut {NoStop}%
\bibitem [{\citenamefont {Perdew}\ \emph {et~al.}(1996)\citenamefont {Perdew},
  \citenamefont {Burke},\ and\ \citenamefont {Ernzerhof}}]{Perdew1996}%
  \BibitemOpen
  \bibfield  {author} {\bibinfo {author} {\bibfnamefont {J.~P.}\ \bibnamefont
  {Perdew}}, \bibinfo {author} {\bibfnamefont {K.}~\bibnamefont {Burke}},\ and\
  \bibinfo {author} {\bibfnamefont {M.}~\bibnamefont {Ernzerhof}},\ }\bibfield
  {title} {\bibinfo {title} {Generalized {{Gradient Approximation Made
  Simple}}},\ }\href {https://doi.org/10.1103/PhysRevLett.77.3865} {\bibfield
  {journal} {\bibinfo  {journal} {Phys. Rev. Lett.}\ }\textbf {\bibinfo
  {volume} {77}},\ \bibinfo {pages} {3865} (\bibinfo {year}
  {1996})}\BibitemShut {NoStop}%
\bibitem [{\citenamefont {Kotani}\ and\ \citenamefont {{van
  Schilfgaarde}}(2002)}]{Kotani2002}%
  \BibitemOpen
  \bibfield  {author} {\bibinfo {author} {\bibfnamefont {T.}~\bibnamefont
  {Kotani}}\ and\ \bibinfo {author} {\bibfnamefont {M.}~\bibnamefont {{van
  Schilfgaarde}}},\ }\bibfield  {title} {\bibinfo {title} {All-electron {GW}
  approximation with the mixed basis expansion based on the full-potential
  {LMTO} method},\ }\href
  {https://doi.org/https://doi.org/10.1016/S0038-1098(02)00028-5} {\bibfield
  {journal} {\bibinfo  {journal} {Solid State Commun.}\ }\textbf {\bibinfo
  {volume} {121}},\ \bibinfo {pages} {461} (\bibinfo {year}
  {2002})}\BibitemShut {NoStop}%
\bibitem [{\citenamefont {van Schilfgaarde}\ \emph {et~al.}(2006)\citenamefont
  {van Schilfgaarde}, \citenamefont {Kotani},\ and\ \citenamefont
  {Faleev}}]{Schilfgaarde2006}%
  \BibitemOpen
  \bibfield  {author} {\bibinfo {author} {\bibfnamefont {M.}~\bibnamefont {van
  Schilfgaarde}}, \bibinfo {author} {\bibfnamefont {T.}~\bibnamefont
  {Kotani}},\ and\ \bibinfo {author} {\bibfnamefont {S.~V.}\ \bibnamefont
  {Faleev}},\ }\bibfield  {title} {\bibinfo {title} {Adequacy of approximations
  in {GW} theory},\ }\href {https://doi.org/10.1103/PhysRevB.74.245125}
  {\bibfield  {journal} {\bibinfo  {journal} {Phys. Rev. B}\ }\textbf {\bibinfo
  {volume} {74}},\ \bibinfo {pages} {245125} (\bibinfo {year}
  {2006})}\BibitemShut {NoStop}%
\bibitem [{\citenamefont {Fuchs}\ \emph {et~al.}(2007)\citenamefont {Fuchs},
  \citenamefont {Furthm\"uller}, \citenamefont {Bechstedt}, \citenamefont
  {Shishkin},\ and\ \citenamefont {Kresse}}]{Fuchs2007}%
  \BibitemOpen
  \bibfield  {author} {\bibinfo {author} {\bibfnamefont {F.}~\bibnamefont
  {Fuchs}}, \bibinfo {author} {\bibfnamefont {J.}~\bibnamefont
  {Furthm\"uller}}, \bibinfo {author} {\bibfnamefont {F.}~\bibnamefont
  {Bechstedt}}, \bibinfo {author} {\bibfnamefont {M.}~\bibnamefont
  {Shishkin}},\ and\ \bibinfo {author} {\bibfnamefont {G.}~\bibnamefont
  {Kresse}},\ }\bibfield  {title} {\bibinfo {title} {Quasiparticle band
  structure based on a generalized {Kohn-Sham} scheme},\ }\href
  {https://doi.org/10.1103/PhysRevB.76.115109} {\bibfield  {journal} {\bibinfo
  {journal} {Phys. Rev. B}\ }\textbf {\bibinfo {volume} {76}},\ \bibinfo
  {pages} {115109} (\bibinfo {year} {2007})}\BibitemShut {NoStop}%
\bibitem [{\citenamefont {Heyd}\ \emph {et~al.}(2003)\citenamefont {Heyd},
  \citenamefont {Scuseria},\ and\ \citenamefont {Ernzerhof}}]{HSE}%
  \BibitemOpen
  \bibfield  {author} {\bibinfo {author} {\bibfnamefont {J.}~\bibnamefont
  {Heyd}}, \bibinfo {author} {\bibfnamefont {G.~E.}\ \bibnamefont {Scuseria}},\
  and\ \bibinfo {author} {\bibfnamefont {M.}~\bibnamefont {Ernzerhof}},\
  }\bibfield  {title} {\bibinfo {title} {Hybrid functionals based on a screened
  {C}oulomb potential},\ }\href {https://doi.org/10.1063/1.1564060} {\bibfield
  {journal} {\bibinfo  {journal} {J. Chem. Phys.}\ }\textbf {\bibinfo {volume}
  {118}},\ \bibinfo {pages} {8207} (\bibinfo {year} {2003})}\BibitemShut
  {NoStop}%
\bibitem [{\citenamefont {Heyd}\ \emph {et~al.}(2006)\citenamefont {Heyd},
  \citenamefont {Scuseria},\ and\ \citenamefont {Ernzerhof}}]{HSEerratum}%
  \BibitemOpen
  \bibfield  {author} {\bibinfo {author} {\bibfnamefont {J.}~\bibnamefont
  {Heyd}}, \bibinfo {author} {\bibfnamefont {G.~E.}\ \bibnamefont {Scuseria}},\
  and\ \bibinfo {author} {\bibfnamefont {M.}~\bibnamefont {Ernzerhof}},\
  }\bibfield  {title} {\bibinfo {title} {Erratum: “{H}ybrid functionals based
  on a screened {C}oulomb potential” [{J. Chem. Phys.} 118, 8207 (2003)]},\
  }\href {https://doi.org/10.1063/1.2204597} {\bibfield  {journal} {\bibinfo
  {journal} {J. Chem. Phys.}\ }\textbf {\bibinfo {volume} {124}},\ \bibinfo
  {pages} {219906} (\bibinfo {year} {2006})}\BibitemShut {NoStop}%
\bibitem [{\citenamefont {Shishkin}\ and\ \citenamefont
  {Kresse}(2006)}]{Shishkin2006}%
  \BibitemOpen
  \bibfield  {author} {\bibinfo {author} {\bibfnamefont {M.}~\bibnamefont
  {Shishkin}}\ and\ \bibinfo {author} {\bibfnamefont {G.}~\bibnamefont
  {Kresse}},\ }\bibfield  {title} {\bibinfo {title} {Implementation and
  performance of the frequency-dependent {GW} method within the {PAW}
  framework},\ }\href {https://doi.org/10.1103/PhysRevB.74.035101} {\bibfield
  {journal} {\bibinfo  {journal} {Phys. Rev. B}\ }\textbf {\bibinfo {volume}
  {74}},\ \bibinfo {pages} {035101} (\bibinfo {year} {2006})}\BibitemShut
  {NoStop}%
\bibitem [{\citenamefont {Hunt}\ \emph {et~al.}(2020)\citenamefont {Hunt},
  \citenamefont {Monserrat}, \citenamefont {Z\'olyomi},\ and\ \citenamefont
  {Drummond}}]{Hunt2020}%
  \BibitemOpen
  \bibfield  {author} {\bibinfo {author} {\bibfnamefont {R.~J.}\ \bibnamefont
  {Hunt}}, \bibinfo {author} {\bibfnamefont {B.}~\bibnamefont {Monserrat}},
  \bibinfo {author} {\bibfnamefont {V.}~\bibnamefont {Z\'olyomi}},\ and\
  \bibinfo {author} {\bibfnamefont {N.~D.}\ \bibnamefont {Drummond}},\
  }\bibfield  {title} {\bibinfo {title} {Diffusion quantum {Monte Carlo} and
  {$GW$} study of the electronic properties of monolayer and bulk hexagonal
  boron nitride},\ }\href {https://doi.org/10.1103/PhysRevB.101.205115}
  {\bibfield  {journal} {\bibinfo  {journal} {Phys. Rev. B}\ }\textbf {\bibinfo
  {volume} {101}},\ \bibinfo {pages} {205115} (\bibinfo {year}
  {2020})}\BibitemShut {NoStop}%
\bibitem [{\citenamefont {Allen}\ and\ \citenamefont
  {Heine}(1976)}]{Allen1976}%
  \BibitemOpen
  \bibfield  {author} {\bibinfo {author} {\bibfnamefont {P.~B.}\ \bibnamefont
  {Allen}}\ and\ \bibinfo {author} {\bibfnamefont {V.}~\bibnamefont {Heine}},\
  }\bibfield  {title} {\bibinfo {title} {Theory of the temperature dependence
  of electronic band structures},\ }\href
  {https://doi.org/10.1088/0022-3719/9/12/013} {\bibfield  {journal} {\bibinfo
  {journal} {J. Phys. C: Solid State Phys.}\ }\textbf {\bibinfo {volume} {9}},\
  \bibinfo {pages} {2305} (\bibinfo {year} {1976})}\BibitemShut {NoStop}%
\bibitem [{\citenamefont {Allen}\ and\ \citenamefont
  {Cardona}(1981)}]{Allen1981}%
  \BibitemOpen
  \bibfield  {author} {\bibinfo {author} {\bibfnamefont {P.~B.}\ \bibnamefont
  {Allen}}\ and\ \bibinfo {author} {\bibfnamefont {M.}~\bibnamefont
  {Cardona}},\ }\bibfield  {title} {\bibinfo {title} {Theory of the temperature
  dependence of the direct gap of germanium},\ }\href
  {https://doi.org/10.1103/PhysRevB.23.1495} {\bibfield  {journal} {\bibinfo
  {journal} {Phys. Rev. B}\ }\textbf {\bibinfo {volume} {23}},\ \bibinfo
  {pages} {1495} (\bibinfo {year} {1981})}\BibitemShut {NoStop}%
\bibitem [{\citenamefont {Tutchton}\ \emph {et~al.}(2018)\citenamefont
  {Tutchton}, \citenamefont {Marchbanks},\ and\ \citenamefont
  {Wu}}]{Tutchton2018}%
  \BibitemOpen
  \bibfield  {author} {\bibinfo {author} {\bibfnamefont {R.}~\bibnamefont
  {Tutchton}}, \bibinfo {author} {\bibfnamefont {C.}~\bibnamefont
  {Marchbanks}},\ and\ \bibinfo {author} {\bibfnamefont {Z.}~\bibnamefont
  {Wu}},\ }\bibfield  {title} {\bibinfo {title} {Structural impact on the
  eigenenergy renormalization for carbon and silicon allotropes and boron
  nitride polymorphs},\ }\href {https://doi.org/10.1103/PhysRevB.97.205104}
  {\bibfield  {journal} {\bibinfo  {journal} {Phys. Rev. B}\ }\textbf {\bibinfo
  {volume} {97}},\ \bibinfo {pages} {205104} (\bibinfo {year}
  {2018})}\BibitemShut {NoStop}%
\bibitem [{\citenamefont {Mishra}\ and\ \citenamefont
  {Bhattacharya}(2019)}]{Mishra2019}%
  \BibitemOpen
  \bibfield  {author} {\bibinfo {author} {\bibfnamefont {H.}~\bibnamefont
  {Mishra}}\ and\ \bibinfo {author} {\bibfnamefont {S.}~\bibnamefont
  {Bhattacharya}},\ }\bibfield  {title} {\bibinfo {title} {Giant exciton-phonon
  coupling and zero-point renormalization in hexagonal monolayer boron
  nitride},\ }\href {https://doi.org/10.1103/PhysRevB.99.165201} {\bibfield
  {journal} {\bibinfo  {journal} {Phys. Rev. B}\ }\textbf {\bibinfo {volume}
  {99}},\ \bibinfo {pages} {165201} (\bibinfo {year} {2019})}\BibitemShut
  {NoStop}%
\bibitem [{\citenamefont {{Lloyd-Williams}}\ and\ \citenamefont
  {Monserrat}(2015)}]{Lloyd-williams2015}%
  \BibitemOpen
  \bibfield  {author} {\bibinfo {author} {\bibfnamefont {J.~H.}\ \bibnamefont
  {{Lloyd-Williams}}}\ and\ \bibinfo {author} {\bibfnamefont {B.}~\bibnamefont
  {Monserrat}},\ }\bibfield  {title} {\bibinfo {title} {Lattice dynamics and
  electron-phonon coupling calculations using nondiagonal supercells},\ }\href
  {https://doi.org/10.1103/PhysRevB.92.184301} {\bibfield  {journal} {\bibinfo
  {journal} {Phys. Rev. B}\ }\textbf {\bibinfo {volume} {92}},\ \bibinfo
  {pages} {184301} (\bibinfo {year} {2015})}\BibitemShut {NoStop}%
\bibitem [{\citenamefont {Monserrat}(2016)}]{Monserrat2016}%
  \BibitemOpen
  \bibfield  {author} {\bibinfo {author} {\bibfnamefont {B.}~\bibnamefont
  {Monserrat}},\ }\bibfield  {title} {\bibinfo {title} {Correlation effects on
  electron-phonon coupling in semiconductors: {{Many-body}} theory along
  thermal lines},\ }\href {https://doi.org/10.1103/PhysRevB.93.100301}
  {\bibfield  {journal} {\bibinfo  {journal} {Phys. Rev. B}\ }\textbf {\bibinfo
  {volume} {93}},\ \bibinfo {pages} {100301} (\bibinfo {year}
  {2016})}\BibitemShut {NoStop}%
\bibitem [{\citenamefont {Monserrat}(2018)}]{Monserrat2018}%
  \BibitemOpen
  \bibfield  {author} {\bibinfo {author} {\bibfnamefont {B.}~\bibnamefont
  {Monserrat}},\ }\bibfield  {title} {\bibinfo {title} {Electron--phonon
  coupling from finite differences},\ }\href
  {https://doi.org/10.1088/1361-648X/aaa737} {\bibfield  {journal} {\bibinfo
  {journal} {J. Phys.: Condens. Matter}\ }\textbf {\bibinfo {volume} {30}},\
  \bibinfo {pages} {083001} (\bibinfo {year} {2018})}\BibitemShut {NoStop}%
\bibitem [{\citenamefont {Towns}\ \emph {et~al.}(2014)\citenamefont {Towns},
  \citenamefont {Cockerill}, \citenamefont {Dahan}, \citenamefont {Foster},
  \citenamefont {Gaither}, \citenamefont {Grimshaw}, \citenamefont {Hazlewood},
  \citenamefont {Lathrop}, \citenamefont {Lifka}, \citenamefont {Peterson},
  \citenamefont {Roskies}, \citenamefont {Scott},\ and\ \citenamefont
  {Wilkins-Diehr}}]{XSEDE}%
  \BibitemOpen
  \bibfield  {author} {\bibinfo {author} {\bibfnamefont {J.}~\bibnamefont
  {Towns}}, \bibinfo {author} {\bibfnamefont {T.}~\bibnamefont {Cockerill}},
  \bibinfo {author} {\bibfnamefont {M.}~\bibnamefont {Dahan}}, \bibinfo
  {author} {\bibfnamefont {I.}~\bibnamefont {Foster}}, \bibinfo {author}
  {\bibfnamefont {K.}~\bibnamefont {Gaither}}, \bibinfo {author} {\bibfnamefont
  {A.}~\bibnamefont {Grimshaw}}, \bibinfo {author} {\bibfnamefont
  {V.}~\bibnamefont {Hazlewood}}, \bibinfo {author} {\bibfnamefont
  {S.}~\bibnamefont {Lathrop}}, \bibinfo {author} {\bibfnamefont
  {D.}~\bibnamefont {Lifka}}, \bibinfo {author} {\bibfnamefont {G.~D.}\
  \bibnamefont {Peterson}}, \bibinfo {author} {\bibfnamefont {R.}~\bibnamefont
  {Roskies}}, \bibinfo {author} {\bibfnamefont {J.~R.}\ \bibnamefont {Scott}},\
  and\ \bibinfo {author} {\bibfnamefont {N.}~\bibnamefont {Wilkins-Diehr}},\
  }\bibfield  {title} {\bibinfo {title} {{XSEDE}: Accelerating scientific
  discovery},\ }\href {https://doi.org/10.1109/MCSE.2014.80} {\bibfield
  {journal} {\bibinfo  {journal} {Comput. Sci. Eng.}\ }\textbf {\bibinfo
  {volume} {16}},\ \bibinfo {pages} {62} (\bibinfo {year} {2014})}\BibitemShut
  {NoStop}%
\end{thebibliography}%

\end{document}